\documentclass[a4paper,11pt]{article}
\pdfoutput=1 
\usepackage{jheppub} 
\usepackage{hyperref}
\usepackage{booktabs}
\usepackage[T1]{fontenc} 
\usepackage{siunitx} 
\usepackage{graphicx}%
\usepackage[dvipsnames]{xcolor}
\usepackage{dcolumn}
\usepackage{bm}
\usepackage[normalem]{ulem}
\usepackage{braket} 
\usepackage{amsmath}
\usepackage{cancel}
\usepackage{slashed}
\usepackage{multicol}
\usepackage[capitalise]{cleveref}
\usepackage{mathtools} 
\usepackage{etoolbox,siunitx}
\robustify\bfseries

\usepackage{subcaption}

\counterwithin*{equation}{section}
\allowdisplaybreaks

\newcommand{\nnl}{\nonumber \\}

\newcommand{\cO}{\mathcal{O}}

\newcommand{\cM}{{\mathcal M}}

\newcommand{\mpl}{M_{\rm Pl}}

\newcommand{\hc}{\rm h.c.}

\begin{document}

\title{
UV completions of scalar-tensor EFTs
}

\author{Edoardo Alviani,}
\author{Adam Falkowski, and}
\author{Panagiotis Marinellis}

\affiliation{Universit\'{e} Paris-Saclay, CNRS/IN2P3, IJCLab, 91405 Orsay, France}

\emailAdd{
alviani@ijclab.in2p3.fr,
adam.falkowski@ijclab.in2p3.fr,
panagiotis.marinellis@ijclab.in2p3.fr
} 

\abstract{
We study models that give rise to scalar-tensor effective field theories (EFTs) at low energies. 
Our framework involves massive particles of spin $S=0, 1/2, 1$ coupled to gravity and to a real massless scalar in the UV. 
Integrating out the massive states leads to a scalar-tensor EFT describing the massless graviton and scalar degrees of freedom. 
Using the on-shell amplitude methods and the spinor-helicity formalism, 
we match the two frameworks at one loop, 
so as to express the EFT Wilson coefficients in terms of the UV masses and coupling.
We explore the space of  
the operators generated in the EFT, including the ones related to the scalar Gauss-Bonnet (SGB) and dynamical Chern-Simons (DCS) gravity theories. 
We demonstrate that, within our setup, the SGB interactions are always generated with shift-symmetry breaking operators. This is in contrast to the DCS case, where there is a unique choice that preserves the shift symmetry in the IR, corresponding to a theory of spin 1/2 fermions and a complex scalar with a Peccei-Quinn global symmetry.
}

\maketitle

\section{Introduction} 
\label{sec:intro}

General relativity (GR) is extremely successful in describing phenomenological aspects of gravitational interactions.  
The theory can be concisely described by the Einstein-Hilbert Lagrangian, 
\begin{align}
\label{eq:INTRO_LGR}
{\cal L}_{\rm GR}  =  &
 \mpl^2 \sqrt{-g} \bigg \{  {1 \over 2} R
 - \Lambda_0^2
 \bigg \} 
, \end{align} 
where the Planck mass is  
$\mpl \equiv (8 \pi G)^{-1/2} \simeq 2.44 \times 10^{18}$~GeV,
$g_{\mu\nu}$ is the metric field describing the massless spin-2 graviton,  
the Ricci scalar is defined in terms of Ricci and Riemann tensors by 
$R = g^{\mu \nu} R_{\mu \nu} =  g^{\mu \nu} R^\alpha{}_{\mu \nu \alpha}$, 
and  $\Lambda_0 \simeq 2 \times 10^{-33}$~eV  is the cosmological constant~\cite{Planck:2018vyg}. 
While there is currently no evidence for gravitational dynamics not described by \cref{eq:INTRO_LGR}, 
 theoretical and experimental studies of GR extensions are continuing. These are motivated by various current problems related to 
 the theory's quantization,
 its high-energy behavior,  dark matter  and dark energy among others, or by purely phenomenological considerations.
The simplest deformation consists in adding new invariants constructed from the Riemann tensor to the Einstein-Hilbert action. 
This can be done within an effective field theory (EFT) framework~\cite{Donoghue:1994dn,Burgess:2003jk,Ruhdorfer:2019qmk}.  
In this so-called GREFT, the interactions beyond GR are organized in a systematic expansion controlled by a scale $\Lambda$ which may be equal to or smaller than the Planck scale controlling GR interactions.
GREFT provides a framework for a fully consistent quantum theory of gravity, which has to be embedded at the scale $\Lambda$ into a more fundamental framework, for example a string theory.

The next simplest deformation is to add to the GR Lagrangian a single  
massless scalar degree of freedom. 
This leads to a class of scalar-tensor theories, where the scalar interacts with gravity in a non-standard way beyond the minimal coupling.  
A scalar that mixes with gravity in vacuum due to its coupling to the Ricci scalar $R$ is very constrained by fifth force measurements. 
Larger couplings are allowed when the scalar is coupled linearly to the Gauss-Bonnet or Chern-Simons (Pontryagin) invariant, leading to the shift-invariant scalar Gauss-Bonnet (SGB)~\cite{Sotiriou:2013qea,Sotiriou:2014pfa} and dynamical Chern-Simons (DCS)~\cite{Alexander:2009tp,Yunes:2009hc} theories: 
\begin{align}  
\label{eq:INTRO_DeltaGR}
\Delta {\cal L}_{\rm GR} \supset &   
 \mpl \sqrt{-g} \phi  
\bigg \{ 
\alpha {\cal R}_{\rm GB}^2
+ \tilde{\alpha} {\cal R}_{\rm CS}^2
\bigg \}  
,   \end{align} 
where $\phi$ is the massless scalar field, ${\cal R}_{\rm GB}^2 \equiv R^{\mu \nu  \rho \sigma}  R_{\mu \nu  \rho \sigma}
- 4R^{\mu \nu}  R_{\mu \nu} + R^2$,
${\cal R}_{\rm CS}^2 \equiv  R^{\mu \nu \alpha \beta} \widetilde R_{\mu \nu \alpha \beta}$, while $\alpha$ and $\tilde \alpha$ are parameters of dimension $[{\rm length}]^2$.
Much as GREFT, the scalar-tensor theories can be organized into a systematic EFT framework~\cite{Serra:2022pzl,Hong:2023zgm} where the shift-symmetric SGB and DCS couplings appear as leading operators in the Lagrangian along with an infinite tower of higher-dimensional operators. 
In particular, the EFT Lagrangian may contain (shift-non-invariant)  couplings of higher powers of $\phi$ to the Gauss-Bonnet and Chern-Simons invariants, often included in the literature by replacing $\phi \to f(\phi)$ in \cref{eq:INTRO_DeltaGR}.

Scalar-tensor theories are currently a focus of an increased experimental scrutiny, providing a useful template for describing departures from GR predictions. 
In particular, the DCS interaction induces parity violation which may be tested by looking at gravitational wave birefringence~\cite{Alexander:2004wk,Okounkova:2021xjv,Yunes:2010yf,Yagi:2017zhb,Alexander:2017jmt} or for parity-violating patterns in the CMB data~\cite{Alexander:2004wk,Lyth:2005jf,Satoh:2010ep}. Tests of the SGB model from X-ray data from binary compact stars~\cite{Yagi:2012gp,Silva:2020acr,Saffer:2021gak} have also been used to provide stringent constraints. 
One spectacular phenomenological impact of the  couplings in \cref{eq:INTRO_DeltaGR} is the prediction of scalar charges (or scalar hair) carried by black holes for both the SGB~\cite{Kanti:1995vq,Maeda:2009uy,Yagi:2011xp,Sotiriou:2013qea,Sotiriou:2014pfa} and DCS~\cite{Grumiller:2007rv,Yunes:2009hc,Alexander:2009tp,Yagi:2012ya} frameworks. 
The presence of scalar hair is especially relevant in the context of gravitational wave emission from inspiral of astrophysical compact objects, since such systems may lose energy faster than predicted in GR by radiating the massless scalar particles~\cite{Yagi:2011xp, Yagi:2012vf,Yagi:2013mbt, Witek:2018dmd,Loutrel:2018ydv,Julie:2019sab,Shiralilou:2021mfl,Julie:2024fwy,Bhattacharyya:2024kxj,Falkowski:2024bgb,Falkowski:2024yuy,Wilson-Gerow:2025xhr}.  
This can be probed using the existing data from the LIGO/VIRGO/KAGRA collaborations~\cite{KAGRA:2022dwb}, as well as from the next generation of gravitational wave interferometers~\cite{LISA:2017pwj,ET:2019dnz,Evans:2021gyd}. 
Current data are able to probe the SGB and DCS interactions with the characteristic scale of the same order as the Schwarzschild radius of the colliding objects, 
that is $|\alpha + i \tilde \alpha|^{1/2} 
\sim (1~\rm km) 
\sim  {1 \over 10^{-10} {\rm eV}}$.  
This is a very low scale for particle physics standards, which makes it relevant to ask the question about possible ultraviolet (UV) completions of the SGB and DCS theories.    

In this paper we study a particular UV completion to scalar-tensor EFTs of gravity. Apart from the  massless graviton and scalar degrees of freedom, it contains massive particles with spin $S=0, 1/2, 1$, which are minimally coupled to gravity and  have general cubic
couplings with the massless scalar.  
Integrating out the heavy particles at one loop generates the interactions in \cref{eq:INTRO_DeltaGR} in the low-energy EFT, together with a tower of other higher-dimensional operators. 
We discuss how the patterns of EFT Wilson coefficients depends on the spectrum and interactions in the UV completion.  

This paper is organized as follows. In \cref{sec:eft} we present the low-energy  Lagrangian for scalar-tensor EFTs. 
We calculate the associated on-shell 3-point amplitudes and use them to construct the tree-level 4-point amplitudes involving scalars and gravitons that are needed to perform the EFT matching. 
In \cref{sec:uv} we describe our UV model. 
We determine the 3-point amplitudes for the interaction of the massless scalar with massive matter and all 4-point amplitudes describing the scattering of the massive particles with the massless scalars and gravitons. We describe our EFT matching strategy in \cref{sec:match} and explicitly calculate the Wilson coefficients of the EFT in terms of the UV couplings of the massive matter to the scalar. 
We highlight the difficulty of generating the purely shift-symmetric version of SGB gravity. 
For the UV completion of the shift-symmetric DCS gravity, the unique predictive UV theory that preserves the shift symmetry of the EFT is the one discussed in Refs.~\cite{Alexander:2022cow,Alexander:2024vav}.  
It features a spin 1/2 massive particle and a complex scalar field endowed with Peccei-Quinn global symmetry $U(1)_{PQ}$. 
We also find that UV theories with only massive $S=1$ particles are not predictive in the context of calculating the Wilson coefficients in \cref{eq:INTRO_DeltaGR}, and we expect this to persist for higher spins as well. 
We conclude in \cref{sec:con}. 
\cref{app:PSI} reviews the details of the phase space integration appearing in our 1-loop matching method.
Finally, \cref{sec:amp} collects the relevant tree- and one-loop UV theory amplitudes that are used in the matching computation. 
  \\ 

\textbf{Conventions:}  Our conventions for the massless spinor helicity formalism are the ones of Ref.~\cite{Dreiner:2008tw}. 
We use the usual angle and square brackets to denote the holomorphic and anti-holomorphic spinors, and we suppress $\sigma$ matrices when sandwiched between spinors. 
The treatment of massive spinors follows   Ref.~\cite{Arkani-Hamed:2017jhn}. 
We define $s,t,u=(p_1+p_{2,3,4})^2$.
We  work with the mostly minus metric $\eta_{\mu \nu} = (1,-1,-1,-1)$, 
and the natural units $\hbar = c = 1$.
The sign convention for the totally anti-symmetric Levi-Civita tensor $\epsilon^{\mu\nu\rho\alpha}$ is 
$\epsilon^{0123} = 1$.  
We use the all-incoming convention for our on-shell amplitudes. 
The Christoffel connection is
$ \Gamma^\mu_{\ \nu \rho} =  
{1 \over 2}  g^{\mu \alpha} \left (
\partial_\rho g_{\alpha \nu} +  \partial_\nu g_{\alpha \rho} - \partial_\alpha g_{\nu \rho} \right ) 
$, 
the Riemann tensor is 
$R^\alpha_{\ \mu \nu \beta}  =   
 \partial_\nu \Gamma^\alpha_{\ \mu \beta} -  \partial_\beta \Gamma^\alpha_{\ \mu \nu}  
 +  \Gamma^\rho_{\ \mu \beta}  \Gamma^\alpha_{\ \rho \nu}
 - \Gamma^\rho_{\ \mu \nu}  \Gamma^\alpha_{\ \rho \beta}  
$.

\section{Low-energy EFT} 
\label{sec:eft}

\subsection{Lagrangian}

We consider a weakly-coupled low-energy EFT of two massless particles: the spin-2 graviton described by the symmetric tensor field $h_{\mu \nu}$, and a spin-0 particle described the real scalar field $\phi$.  
In this paper we ignore other degrees of freedom that may be present at low energies, in particular the stable SM particles.\footnote{%
Gravitational EFTs are often used at low enough energy scales, where massive unstable SM particles can be integrated out.} 
The theory is described by the Lagrangian organized as 
\begin{align}
\label{eq:EFT_Left}
{\cal L}_{\rm EFT} =  & 
 \sqrt{-g} \bigg \{  
{\cal L}_0 + 
\sum_{D \geq 5}  {\cal L}^{(D)}
 \bigg \}  
. \end{align} 
The Lagrangian respects general coordinate invariance with the metric related to the graviton field by 
$g_{\mu\nu} = \eta_{\mu \nu} 
+ {2 \over \mpl} h_{\mu \nu}$, 
where $\eta$ is the Minkowski metric. 
The first term, ${\cal L}_0$, contains operators of dimension $D\leq 4$. 
This includes the Einstein-Hilbert term describing the GR dynamics of the graviton, the kinetic term of the scalar and its minimal coupling to gravity, and  scalar self-interactions: 
\begin{align}
\label{eq:EFT_L0}
{\cal L}_0  =  &
 {\mpl^2 \over 2}  R 
+  {1 \over 2} \partial_\mu \phi 
\partial^\mu \phi  
- {C_3 \over 3!} \phi^3 
- {C_4 \over 4!} \phi^4 
. \end{align} 
For the sake of this paper, these are all terms in the EFT Lagrangian $D \leq 4$.  
In particular,  we ignore the small cosmological constant so that we can work with a quantum field theory on the flat Minkowski background. Furthermore,  \cref{eq:EFT_L0} does not contain the dimension-2 scalar mass term $\phi^2$, in line with our starting assumption that the scalar is massless.
The rationale could be the shift symmetry $\phi \to \phi + \alpha$, which forbids 
the scalar mass term, in which case also $C_3 = C_4 =0$. 
Without the shift symmetry, the masslessness of $\phi$  must be enforced by fine-tuning. 
Other low-dimensional operators allowed by general coordinate invariance are either total derivatives (the Chern-Simons term 
or the Gauss-Bonnet invariants) 
or can be eliminated using field redefinitions 
($R^{\mu \nu}  R_{\mu \nu}$, 
$R^2$). 
For the latter reason we can also ignore scalar couplings to the Ricci tensor (
$\phi R$, 
$\phi^2 R$, 
$\phi \partial^\mu \partial^\nu R_{\mu \nu}$, etc.)\footnote{In this paper we focus on the scalar-tensor sector and we ignore the operators involving matter fields that are generated by such field redefinitions.  }

Furthermore, ${\cal L}^{(D)}$ in \cref{eq:EFT_Left} contains dimension-$D$ operators constructed out of the scalar and metric fields. 
The Wilson coefficients in ${\cal L}^{(D)}$ are dimensionful and scale as $C \sim \cO(\Lambda^{4-D})$,  where $\Lambda$ is the scale controlling the EFT expansion.
Ultimately, $\Lambda$ will be identified with the mass scale of heavy particle in the UV completion of our EFT. 
From now on we will always assume $\Lambda \ll \mpl$, such that the UV completion can be a weakly coupled quantum field theory.

Beginning with dimension five, 
a complete basis of operators at this order reads 
\begin{align}  
\label{eq:EFT_L5}
{\cal L}^{(5)} = &   
{1 \over 2 } \phi  
\bigg \{ 
C_{\phi \rm GB} {\cal R}_{\rm GB}^2
+ C_{\phi \rm CS} {\cal R}_{\rm CS}^2
\bigg \}  
- {C_5 \over 5!} \phi^5 
,   \end{align} 
where ${\cal R}_{\rm GB}^2 \equiv R^{\mu \nu  \rho \sigma}  R_{\mu \nu  \rho \sigma}
- 4R^{\mu \nu}  R_{\mu \nu} + R^2$ is the Gauss-Bonnet invariant, and 
${\cal R}_{\rm CS}^2 \equiv  R^{\mu \nu \alpha \beta} \widetilde R_{\mu \nu \alpha \beta}$ is the Chern-Simons invariant. 
The part containing the curly bracket describes the shift-symmetric SGB and DCS couplings, which are the main focus of this paper.\footnote{%
Compared to the more standard notation in \cref{eq:INTRO_DeltaGR}, 
we redefined 
$C_{\phi \rm GB} = 2 \mpl \alpha$,
$C_{\phi  \rm CS} = 2 \mpl \tilde \alpha$. 
The present notation is more convenient for the matching with UV completions we consider later. 
}
The goal of this paper is to  investigate how the Wilson coefficients $C_{\phi \rm GB}$ and $C_{\phi \rm CS}$ are related to the parameters of simple UV completions of our EFT. 
If only $C_{\phi \rm GB}$ ($C_{\phi \rm CS}$) is non-zero then  $\phi$ can be assigned even (odd) parity; if both are non-zero then parity is broken. 
Note that these operators respect the shift symmetry, which is due to the fact that  ${\cal R}_{\rm GB}^2$ and ${\cal R}_{\rm CS}^2$  are both total derivatives. 
The last term in \cref{eq:EFT_L5} does not respect the shift symmetry. It will play no role in our discussion and is only included for completeness.

Moving to dimension six, 
a complete basis of operators at this order reads
\begin{align}  
\label{eq:EFT_L6}
{\cal L}^{(6)} = &  
{C_{R^3} \over 3! }  
{\cal R}^3 
+{\tilde C_{R^3} \over 3! }
 \widetilde{\cal R}^3
+ {1 \over 4 } \phi^2  
\bigg \{ 
   C_{\phi^2 \rm GB} 
   {\cal R}_{\rm GB}^2 
+ C_{\phi^2 \rm CS} 
{\cal R}_{\rm CS}^2 
\bigg \}  
- {C_6 \over 6!} \phi^6 
,   \end{align} 
where ${\cal R}^3  \equiv 
C_{\mu \nu  \rho \sigma} C^{\rho \sigma \alpha \beta} C_{\alpha \beta}{}^{\mu \nu }$,  $\widetilde{\cal R}^3  \equiv 
C_{\mu \nu  \rho \sigma} C^{\rho \sigma \alpha \beta} \widetilde C_{\alpha \beta}{}^{\mu \nu }$, 
and the Weyl tensor 
$C_{\mu \nu  \rho \sigma}$ is the traceless part of the Riemann tensor. 
The first line describes the leading corrections to GR, with the first term being parity-conserving and the second parity-breaking. 
The remainder of dimension-6 operators does not respect the  shift symmetry. 
From this perspective, we will be particularly interested in the scalar-squared-Gauss-Bonnet and scalar-squared-Chern-Simons couplings. Some of the UV completions we will later consider violate the shift symmetry, and 
a non-zero value  of $C_{\phi^2 \rm GB}$ or $C_{\phi^2 \rm CS}$ is a  convenient indicator of the absence of that symmetry.  
The $\phi^6$ operator will play no role in our following discussion.

For the sake of this paper we will not be interested in the operator bases at larger dimensions. 
We only highlight a subset of such operators that will be relevant for the following discussion. 
First, at dimension seven, we have the operators 
\begin{align}  
\label{eq:EFT_L7}
{\cal L}^{(7)} \supset 
{  \phi  \over 3!  }     \bigg \{ 
 C_{\phi \rm R^3} {\cal R}^3 
+\widetilde C_{\phi \rm R^3} 
\widetilde{\cal R}^3
 \bigg \} 
,  \end{align} 
which can serve as another indicator of the absence of shift symmetry in the EFT. 
We will also discuss the dimension-8 operator:  
\begin{align}
\label{eq:EFT_L8}
{\cal L}^{(8)} \supset 
{C_4' \over 2} 
(\partial_\mu \phi \partial^\mu \phi)
(\partial_\nu \phi \partial^\nu \phi)
. \end{align}
This is the subleading correction to the 4-scalar self-interaction in this EFT (there is no 4-scalar operator at dimension six). 
In a shift invariant theory, where $C_4$ in \cref{eq:EFT_L0} is forbidden, this becomes the leading self-interaction. 

\subsection{On-shell amplitudes}

Given the Lagrangian in \cref{eq:EFT_Left} we can calculate scattering amplitudes in the EFT. 
These will later be compared to analogous amplitudes in the UV completion, which will allow us to determine the dependence of the EFT Wilson coefficients on the UV particles' masses and couplings. 
Our primary goal will be to match the SGB and DCS Wilson coefficients in \cref{eq:EFT_L5}.  
Nevertheless, we are also interested in a more global picture regarding the pattern of higher-dimensional EFT operators generated by simple UV completions of this EFT. 

The SGB and DCS couplings contribute already at the level of 3-particle  amplitudes. 
The complete list of on-shell 3-point all-incoming amplitudes in this EFT is as follows: 
\begin{align}
\label{eq:EFT_Mhhh}
\cM[1_h^- 2_h^- 3_h^+] = &  
 -{1 \over \mpl}   
 {  \langle 12 \rangle^6 
 \over \langle 13 \rangle^2 \langle 23 \rangle^2 } 
, \nnl 
\cM[1_h^+2_h^+ 3_h^-] = &   
- {1 \over \mpl}   {
[12]^6 \over [13]^2 [23]^2 }
, \nnl 
\cM[1_h^-  2_h^- 3_h^-] = & 
 {C_{R^3} + i \widetilde C_{R^3}  \over \mpl^3 }  
\langle 12 \rangle^2
\langle 13 \rangle^2 
\langle 23 \rangle^2  
, \nnl 
\cM[1_h^+  2_h^+ 3_h^+] = & 
 {C_{R^3} - i \widetilde C_{R^3}  \over \mpl^3  }  
  [12]^2 [13]^2 [23]^2 
, \end{align} 
\begin{align}
\label{eq:EFT_Mhhphi}
 \cM[1_h^- 2_h^- 3_\phi ]  =  & 
 {C_{\phi \rm GB} + i C_{\phi \rm CS}  \over \mpl^2} 
 \langle 12 \rangle^4   
 , \nnl 
\cM[1_h^+ 2_h^+ 3_\phi ]  =  & 
 {C_{\phi \rm GB}  
 -  i C_{\phi \rm CS}
 \over \mpl^2} [12]^4  
 , \end{align}  
\begin{align}
\label{eq:EFT_Mhphiphi}
\cM[1_h^- 2_\phi 3_\phi] = & 
- {1 \over \mpl}   {
\langle 12 \rangle^2 
\langle 13 \rangle^2 \over  
\langle 23 \rangle^2} 
, \nnl  
 \cM[1_h^+2_\phi 3_\phi] = & 
- {1 \over \mpl}   {[12]^2 [13]^2 \over  [23]^2}
. \end{align}
\begin{align}
 \cM[1_\phi 2_\phi 3_\phi] = &
 -  C_3 
. \end{align}

In the next section we will perform one-loop matching of the EFT Wilson coefficients to UV completions using the fully on-shell approach and unitarity bootstrap. 
This is very convenient in gravitational theories where the off-shell baggage is particularly heavy. 
However,  matching at the level of 3-point on-shell amplitudes is not possible, because all kinematic invariants vanish for massless external particles, and the analytic structure cannot be defined. 
For this reason we turn to 4-point amplitudes. 
Below we list the relevant tree-level on-shell 4-point amplitudes in the EFT, which will be later matched to analogous amplitudes in the UV completions at one loop. 
We are primarily interested in the EFT amplitudes depending on the shift-symmetric SGB and DCS couplings. 
These are particular 3- and 1-graviton amplitudes:  
\begin{align}
\label{eq:EFT_Mhhhphi}
 \cM[1_h^- 2_h^-3_h^+ 4_\phi ]  =  &
- {C_{\phi \rm GB} 
+ i C_{\phi \rm CS}
\over \mpl^3 }  
{\langle 12 \rangle^4         [3| p_1 p_2 |3]^2  \over  s t u }  
 + {\rm contact} 
,  \end{align}  
\begin{align}
\label{eq:EFT_Mhphiphiphi} 
\cM[1_h^- 2_\phi 3_\phi 4_\phi ]  =  &
\langle 1| p_2  p_3  |1\rangle^2   \bigg \{
{ C_3 \over s t u } 
+{ C_{\phi \rm GB} 
+ i C_{\phi \rm CS}
\over \mpl^3}  
 \bigg [ {1 \over s} + {1 \over t} + {1 \over u} \bigg ] 
 \bigg \} 
 + {\rm contact} 
 ,  \end{align}   
as well as their parity conjugates which can be easily obtained by CPT and crossing. 
The amplitudes in \cref{eq:EFT_Mhhhphi,eq:EFT_Mhphiphiphi} are displayed up to contact terms, which are sensitive to operators of dimension larger than six and will be of no interest to us.
On the other hand, the 
$C_{\phi \rm GB}$ and $C_{\phi \rm CS}$ couplings contribute to the residues of simple kinematic poles in the 4-point amplitude, which can be unambiguously identified in the matching procedure. 

In addition, we will be interested in the same-helicity 2-graviton-2-scalar amplitude, which is sensitive to dimension-6 operators: 
\begin{align}  
\label{eq:EFT_Mhhphiphi}
\cM[1_h^- 2_h^- 3_\phi 4_\phi ]   = & 
 \langle 1 2 \rangle^4   
 \bigg \{ 
 - {C_{R^3} + i  \widetilde C_{R^3}  \over \mpl^4 }   {t u  \over s} 
 + 
{C_{\phi^2 \rm GB} + 
i C_{\phi^2 \rm CS}\over \mpl^2} 
\bigg \} 
 + \dots 
  . \end{align} 
Here we have explicitly displayed the contact term corresponding to the 3rd and 4th  dimension-6 operator in \cref{eq:EFT_L6}. 
Matching to the 2-graviton-2-scalar amplitude calculated within a UV completion will allows us to determine the corresponding Wilson coefficients 
$C_{\phi^2 \rm GB}$ and $C_{\phi^2 \rm CS}$. 
The ${\cal R}^3$ operators beyond GR affect the $s$-channel residue in this amplitude. 
Matching of the corresponding Wilson coefficients was performed using the 4-graviton amplitude in Refs.~\cite{Bern:2021ppb,Alviani:2024sxx}. 
Our calculation in the next section provides an alternative way to match $C_{R^3}$ and $\widetilde C_{R^3}$ and offers a useful cross-check. 
The dots stand for contact terms probing dimension-8 and higher operators, as well as
$\cO(\mpl^{-4})$ contact terms. 
The latter are subleading to the contact terms displayed in  \cref{eq:EFT_Mhhphiphi}, which are $\cO(\Lambda^{-2} \mpl^2)$ in the EFT counting, given our assumption $\Lambda \ll \mpl$.  

\section{UV completion} 
\label{sec:uv}

We assume that our EFT of the graviton and massless scalar is UV completed by massive particles,  in the following referred to as heavy matter. 
These can be proxies for the unstable Standard Model particles, or represent unknown hidden sector states. 
Since the UV theory, much as the EFT, is coupled to gravity, it is necessarily an effective theory with an upper bound on the validity (cutoff) scale of order $\mpl$. 
We denote the cutoff scale of the UV theory as $M_*$, and assume $m \ll M_* \lesssim \mpl$. 

To be specific, we introduce a complex spin-0 scalar $\Phi$ with mass $m_\Phi$, a  spin-1/2 Dirac fermion $\Psi$ with mass $m_\Psi$, and a complex spin-1 vector $V_\mu$ with mass $m_V$. 
For spin 1, the mass could be generated  starting from a gauge-invariant theory through a vacuum expectation value of some scalar $H$ charged under $V_\mu$. 
But the details of how $m_V$ appears are unimportant for our discussion and we can think of $H$ as a part of the more fundamental theory at $E \gtrsim M_*$.  

We assume a weakly coupled  UV theory and that the heavy matter particles have a  minimal coupling to gravity. 
This can be described by the 3-point on-shell amplitudes~\cite{Arkani-Hamed:2017jhn}: 
\begin{align}
\label{eq:UV_M3h}
\cM[1_h^-  2_X 3_{\bar X} ]  =  &
- { \langle 1 |  p_2 |\zeta]^2 \over \mpl [ 1 \zeta]^2 } 
{ [ \boldsymbol{2} \boldsymbol{3}]^{2S} 
\over m_X^{2S}} 
, \nnl 
\cM[1_h^+ 2_X 3_{\bar X} ]  =  &
- {[1|  p_2|\zeta \rangle^2 \over \mpl \langle 1  \zeta \rangle^2 } 
{ \langle \boldsymbol{2} \boldsymbol{3} \rangle^{2S} 
\over m_X^{2S}} 
,  \end{align}  
where $S=0$ for $X=\Phi$, $S=1/2$ for $X=\Psi$, and $S=1$ for $X=V$. 

We also assume the massless scalar couples to the heavy matter fields. 
The cubic couplings are described by the following Lagrangian:
\begin{align} 
\label{eq:UV_LUV} 
{\cal L}_{\rm UV} \supset  &
- \phi \bigg \{ 
 y_0  m_\Phi  \phi |\Phi|^2 
 +  y_s \phi (\bar \Psi \Psi )
 -  i y_p  \phi (\bar \Psi \gamma^5 \Psi ) 
 \nnl &  
+ { 1 \over M_*} 
\bigg [  
y_d  V_{\mu \nu} \bar V^{\mu \nu} 
+ y_a  \widetilde V_{\mu \nu} \bar V^{\mu \nu} 
- y_m m^2  V_{\mu} \bar V^{ \mu}
\bigg ] 
\bigg \} 
. \end{align} 
The scalar couplings to spin-0  and spin-1/2 matter are renormalizable. 
In the absence of the spin-1 matter, the UV completion is renormalizable in the limit $\mpl\to \infty$, and therefore it can be a valid effective theory all the way to the Planck scale. 
On the other hand, the scalar couplings to spin-1 matter are non-renormalizable and controlled by the cutoff scale $M_*$. 
Around that scale, the UV completion itself must be UV completed into a more fundamental theory (for example with scalars and fermions charged under $V_\mu$) generating the dimension-5 interactions in \cref{eq:UV_LUV} in the low-energy approximation.   
Note that the spin-1 coupling proportional to $y_m$ could arise from a gauge-invariant operator ${\cal L} \supset - {y_m \over M_*} |D_\mu H|^2$, which justifies the parametrization in \cref{eq:UV_LUV}. 

The on-shell 3-point amplitudes following from \cref{eq:UV_LUV} are 
  \begin{align}
  \label{eq:UV_M3phi}
  \cM[1_\phi  2_\Phi 3_{\bar \Phi} ]  =  &
  - y_0  m_\Phi  
,\nnl 
\cM[1_\phi  2_\Psi 3_{\bar \Psi} ]  =  & 
- y_s \big [ \langle \boldsymbol{2} \boldsymbol{3} \rangle +   [ \boldsymbol{2} \boldsymbol{3}] \big ] 
- i y_p  \big [ 
\langle \boldsymbol{2} \boldsymbol{3} \rangle
-   [ \boldsymbol{2} \boldsymbol{3}] \big ] 
,\nnl 
\cM[1_\phi  2_V 3_{\bar V} ]  =  &  - {1 \over M_*}  \bigg \{  
 y_d \big [  
 \langle \boldsymbol{2} \boldsymbol{3} \rangle^2
 +   [ \boldsymbol{2} \boldsymbol{3}]^2  \big ] 
+  i y_a  \big [  
\langle \boldsymbol{2} \boldsymbol{3} \rangle^2 -  [ \boldsymbol{2} \boldsymbol{3}]^2  \big ] 
+  y_m  \langle \boldsymbol{2} \boldsymbol{3} \rangle
[ \boldsymbol{2}\boldsymbol{3}]
\bigg \} 
.   \end{align}   

\cref{eq:UV_M3h,eq:UV_M3phi} together with \cref{eq:EFT_Mhhh,eq:EFT_Mhphiphi} are the basic building blocks from which all the amplitudes in this UV theory can be constructed using the unitarity bootstrap. 
In the following section we will construct one-loop amplitudes for scattering of the massless degrees of freedom, with heavy matter particles propagating in the loop. 
This will allow us to match the UV theory to the EFT from \cref{sec:eft}. 
In this construction we will need the tree-level 4-point amplitudes describing scattering of massless particles into heavy matter pairs. 
For the scalar matter the relevant amplitudes are listed below: 
\begin{align}
\label{eq:UV_MhhPhiPhi}
\cM[1_h^- 2_h^+ 3_\Phi 4_{\bar \Phi} ]  =  &
{\langle 1| p_3  |2]^4 \over \mpl^2 s (t-m_\Phi^2)(u-m_\Phi^2) } 
, \nnl 
\cM[1_h^- 2_h^- 3_\Phi 4_{\bar \Phi} ]  =  &
{m_\Phi^4 \langle12\rangle^4 \over \mpl^2 s (t-m_\Phi^2)(u-m_\Phi^2) } 
, \end{align}  
\begin{align}
\label{eq:UV_MhphiPhiPhi}
\cM[1_h^- 2_\phi 3_\Phi 4_{\bar \Phi}  ]  =  &
{ y_0 m_\Phi \langle 1| p_2  p_3 | 1\rangle^2  \over \mpl s (t-m_\Phi^2)(u-m_\Phi^2) }
, \end{align}  
\begin{align}  
\label{eq:UV_MphiphiPhiPhi}
\cM[1_\phi 2_\phi 3_\Phi 4_{\bar \Phi} ] =  & 
{ (t  - m_\Phi^2) (u-m_\Phi^2)  \over \mpl^2 s }  
+ { y_0^2   m_\Phi^2  s  \over (t  - m_\Phi^2) (u-m_\Phi^2)  }   
-  c_\Phi 
.  \end{align}  
The dimensionless coupling $c_\Phi$ corresponds to a contact term in the on-shell language, which can be interpreted as coming from a $\phi^2 |\Phi|^2$ interaction in $\cal{L}_{\rm UV}$. 
We include it here for completeness and will comment on its effect for the EFT matching in the next section. 
For the spin-1/2 and spin-1 matter the relevant amplitudes are more lengthy and have been relegated to \cref{sec:amp}. 
These amplitudes also include contact terms, corresponding to the freedom of adding  to ${\cal L}_{\rm UV}$ a coupling of two real scalars to the $\bar{\Psi}\Psi$, $\bar{\Psi}\gamma^5\Psi$, $V_{\mu \nu} \bar V^{\mu \nu}$, 
  $\widetilde V_{\mu \nu} \bar V^{\mu \nu}$ and $V_{\mu} \bar V^{ \mu}$ bilinears. 
  Their coefficients are denoted $c_\Psi$, $\tilde{c}_\Psi$, $c_V$, $\tilde{c}_V$, $\hat{c}_V$, respectively.
 
\section{Matching} 
\label{sec:match}

In this section we perform the matching between the parameters of the UV theories defined in \cref{sec:uv} and the Wilson coefficients in the low-energy EFT Lagrangian in \cref{eq:EFT_Left}. 
Our method closely follows the one described in Ref.~\cite{Alviani:2024sxx}. 
Below we describe in some detail the method as applied to the case of the scalar matter UV completion.  
For the spin-1/2 and spin 1 matter the calculations are analogous, and we only present the final results for the Wilson coefficients, while relegating expressions of the full loop amplitudes to~\cref{sec:amp}. 

\subsection{Procedure}

The shift-symmetric SGB and DCS couplings can be matched by calculating the $\cM[1_h^- 2_\phi 3_\phi 4_\phi]$ in the UV theory, expanding the result for large masses of the matter particle, and comparing with the EFT result in \cref{eq:EFT_Mhphiphiphi} to fix $C_{\phi \rm GB}$ and $C_{\phi \rm CS}$\footnote{%
Alternatively, 
we could calculate $\cM[1_h^- 2_h^- 3_h^+ 4_\phi]$  and compare with \cref{eq:EFT_Mhhhphi}, which of course leads to the same result for $C_{\phi \rm GB}$ and $C_{\phi \rm CS}$.  }.
We apply this procedure at one loop to the spin-0 matter UV completion. 

Let us denote the $s$-channel discontinuity of the UV amplitude by 
${\rm Disc}_s \cM[1_h^- 2_\phi 3_\phi 4_\phi] \equiv \theta(s - 4m_\Phi^2) D_s$, 
where ${\rm Disc}_s f(s,t) \equiv f(s+i\epsilon,t)-f(s-i\epsilon,t)$. 
Unitarity dictates that, at one loop, 
 \begin{align} 
D_s = &  i \int d \Pi_{XY} \cM[1_h^- 2_\phi (-Y)_\Phi (-X)_{\bar \Phi}] \cM [3_\phi 4_\phi X_\Phi Y_{\bar \Phi}] 
 , \end{align} 
where   $d\Pi_{XY}$ is the two-body phases space element for the intermediate spin-0 matter pair with momenta $p_X$ and $p_Y$ and 
the tree-level 4-point amplitudes under the integral are those in \cref{eq:UV_MhphiPhiPhi,eq:UV_MphiphiPhiPhi}. 

The $s$-channel discontinuity at the linear  order in $y_0$ is
 \begin{align} 
\label{eq:MATCH_DsMhphiphiphi}
D_s = &  
 i {y_0 m_\Phi  \langle 12 \rangle^2  \over  \mpl s}   \int d \Pi_{XY} 
 { \langle 1|p_X|2]^2   
 \over (2 p_1 p_X)(2 p_2 p_X)}
 \bigg \{ 
 {(2 p_3 p_X) (2 p_4 p_X) \over \mpl^2 s } 
-     c_\Phi   
+ { y_0^2 m_\Phi^2 
  \over    (2 p_3 p_X ) (2 p_4 p_X)  }    
  \bigg \}  
. \end{align}  
The phase space integral can be handled by changing the variable as 
\begin{align}
p_X=\alpha p_1+(1-\alpha) p_2-\sqrt{\alpha(1-\alpha)-\frac{m_\Phi^2}{s}}[zq+z^{-1} \bar q] 
, \end{align}
where 
$z = e^{i \phi}$, 
$q \sigma = |2\rangle [1|$, $\bar q \sigma = |1\rangle [2|$, and $p_Y = p_1+p_2-p_X$.
The integral over the 2-body phase space  is traded for integration over the $\alpha$ and $z$ variables:  
\begin{align}
\label{eq:MATCH_phasespace}
\int \text{d}\Pi_{XY}=
\int_{\alpha_-}^{\alpha^+}\frac{\text{d}\alpha}{8\pi}
\int_{|z|=1}\frac{\text{d}z}{2\pi i z}
, \end{align}
where 
$\alpha_\pm=\frac{1\pm\sqrt{1-\frac{4m_\Phi^2}{s}}}{2}$. 
After this change of variables,  the discontinuity in \cref{eq:MATCH_DsMhphiphiphi}
can be calculated analytically. 
While the result of the integral is not particularly illuminating by itself, it can be recast as the $s$-channel discontinuity of a linear combination of the scalar integrals defined in \cref{eq:PSI_scalar-integrals}:  
 \begin{align}  
\label{eq:MATCH_DsMhphiphiphi-result}
 D_s   = &   
 \langle 1|p_2 p_3|1 \rangle^2 
 m_\Phi^3 
 {\rm Disc}_s  \bigg \{  
 {y_0  \over \mpl^3 s^3} 
\bigg (  2 m_\Phi^2 I_\triangle^s 
+ {10 m_\Phi^2 - s \over 6 m_\Phi^2} I_\circ^s \bigg )
 \nnl & 
+  {y_0^3  \over \mpl  s t u}  \bigg (
 \bigg [ 2 m_\Phi^2  + {s t \over u} \bigg ] I_{\Box}^{st} 
 +  \bigg [ 2 m_\Phi^2  + {s u \over t} \bigg ] I_{\Box}^{su}  
+ 2 \bigg [ 1+ {t \over u} + {u \over t} \bigg ]    I_\triangle^s  
\bigg ) 
    \bigg \}  
 . \end{align}  
Note that the phase space integral of the piece  proportional to $c_\Phi$ vanishes, therefore  the contact term ambiguity does not affect this particular   computation. 
  
The discontinuities in the $t$ and $u$ channel are obtained from \cref{eq:MATCH_DsMhphiphiphi-result} by 
$2 \leftrightarrow 3$ and  
$2 \leftrightarrow 4$, respectively. 
With all the discontinuities at hand, we can reconstruct the amplitude up to the rational terms and tadpoles: 
\begin{align} 
\label{eq:MATCH_Mhphiphiphi-full}
\cM[1_h^- 2_\phi 3_\phi 4_\phi]  = &   \langle 1|p_2 p_3|1 \rangle^2  
\bigg \{ 
 {y_0  m_\Phi^3  \over \mpl^3}  
 \bigg ( 
 {2 m_\Phi^2  \over s^3}  I_\triangle^s 
 + {10 m_\Phi^2 - s \over 6 m_\Phi^2 s^3} I_\circ^s
 + (s \to t) + (s \to u) 
 \bigg )
 \nnl & 
 +  {y_0^3 m_\Phi^3  \over \mpl s t u }    \bigg \{ 
 \bigg [ 2 m_\Phi^2  
 + {s t \over u} \bigg ] I_{\Box}^{st} 
 +  \bigg [ 2 m_\Phi^2  
 + {s u \over t} \bigg ] I_{\Box}^{su}  
  +  \bigg [ 2 m_\Phi^2 
  + {t u \over s} \bigg ] I_{\Box}^{tu}   
  \nnl & 
+ 2 \bigg [ 1+ {t \over u} + {u \over t} \bigg ]    I_\triangle^s  
+ 2 \bigg [ 1+ {s \over u} + {u \over s} \bigg ]    I_\triangle^t  
+ 2 \bigg [ 1+ {t \over s} + {s \over t} \bigg ]    I_\triangle^u  
 \bigg ) 
+  c_1 I_1 + {\cal R}
 \bigg \} 
 . \end{align}  
To make a connection to the EFT of~\cref{sec:eft} we need to consider  \cref{eq:MATCH_Mhphiphiphi-full} in the low-energy  regime where all the Mandelstam invariants are below the mass scale of the matter particle,  
$s,t,u \sim E^2 \ll m_\Phi^2$. 
In this regime we can expand the amplitude in powers of $E/m_\Phi$. 
Let us first focus on the negative powers of $E/m_\Phi$ appearing in this expansion.
After adjusting the tadpole coefficient $c_1 = {y_0 (s- 10 m_\Phi^2 ) \over 6 \mpl^3 s^3 m_\Phi} + (s \to t,u)$
so as to cancel non-local UV divergent terms in the amplitude (which can only be canceled this way), we obtain 
   \begin{align} 
\cM[1_h^- 2_\phi 3_\phi 4_\phi]  = &  
 \langle 1|p_2 p_3|1\rangle^2 \bigg \{ 
  {y_0  \over 6 \pi^2 \mpl^3 }  
  \bigg ( 
- m_\Phi^3 
\bigg [   {1 \over s^3}
  +   {1 \over t^3} 
 +    {1 \over  u^3} \bigg ]
 +  { 13 m_\Phi  \over 96} 
\bigg [   {1 \over s^2}
  +   {1 \over t^2} 
 +    {1 \over  u^2} \bigg ]
 \bigg ) 
 \nnl & 
 + {m_\Phi y_0^3 \over 
 16 \pi^2 \mpl s t u }
 + {\cal R} 
 + \cO(m_\Phi^{-1} ) 
\bigg \} .
\end{align} 
The first term in the curly bracket contains non-local poles and therefore it has to  be canceled by the rational term. 
On the other hand, the second term proportional to $y_0^3$ is legal, and an analogous expression is allowed to appear in the rational term. 
We write  
${\cal R} = 
{y_0  \over 6 \pi^2 \mpl^3 }  
  \bigg ( 
 { m_\Phi^3 \over s^3}
-  { 13 m_\Phi  \over 96 s^2} 
\bigg ) +  
 {m_\Phi y_0^3 
 (\delta_{\cal R}-1) 
 \over 
 16 \pi^2 \mpl s t u }
$, 
where the rational term ambiguity is parametrized by 
$\delta_{\cal R}$. 
The crucial point is that the {\em entire} ambiguity is encapsulated by $\delta_{\cal R}$. 
In particular, we are not allowed to introduce rational terms scaling with inverse powers of $m_\Phi$, as then the UV theory would not be well defined in the limit of $m_\Phi \to 0$. 

Having fixed the tadpole and rational terms, we can continue the low-energy expansion of the amplitude in \cref{eq:MATCH_Mhphiphiphi-full} to higher orders in $(E/m_\Phi)$. 
We find 
   \begin{align} 
\label{eq:MATCH_Mhphiphiphi-expanded}
\cM[1_h^- 2_\phi 3_\phi 4_\phi]  = &  
 \langle 1|p_2 p_3|1\rangle^2
 \bigg \{ 
{\color{blue}
{m_\Phi y_0^3 
\delta_{\cal R} 
\over 
 16 \pi^2 \mpl s t u }
}
- {\color{red} 
 {y_0  \over  1440 \pi^2 \mpl^3 m_\Phi}  \bigg [ {1 \over s} + {1 \over t}  +  {1 \over u}  \bigg ] }
 \nnl &  
-  {\color{brown}  
{y_0 \over  8960 \pi^2 \mpl^3 m_\Phi^3} 
}  
+ {\color{brown}  
{ y_0^3 \over 40320 \pi^2 \mpl m_\Phi^5} 
}
+  {\color{brown}  
 \cO(m_\Phi^{-7}) }
 \bigg \} 
 . \end{align}  
Matching it with the EFT expression in \cref{eq:EFT_Mhphiphiphi}, 
the blue term, in principle,  allows one to determine the Wilson coefficient $C_3$ controlling the massless scalar cubic coupling. 
That is however not possible using our method due to the rational term ambiguity.
We can only conclude, in line with naive dimensional analysis, that magnitudes of $C_3$ smaller than $\sim {y_0^3  m_\phi \over 16 \pi^2}$ would not be natural. 
More generally, all terms scaling with non-negative powers of the heavy particles masses are subject to the  rational term ambiguity, and therefore with our methods we cannot match EFT Wilson coefficients of operators with dimension four  or less. 

 Moving to the red term, it has the same form as  the graviton pole terms in the EFT amplitude \cref{eq:EFT_Mhphiphiphi} proportional to the SGB and DCS Wilson coefficients $C_{\phi \rm GB}$ and $C_{\phi \rm CS}$. 
 We can match 
    \begin{align} 
C_{\phi \rm GB} = & 
-  {y_0  \over 1440  \pi^2   m_\Phi}  
, \nnl 
C_{\phi \rm CS}  = & 
   0 
.    \end{align}  
Finally, the brown terms in 
\cref{eq:MATCH_Mhphiphiphi-expanded} match to local EFT operators of dimension seven and larger. 

Our next goal is to match the dimension-6 Wilson coefficients 
$C_{\phi^2 \rm GB}$ 
and 
$C_{\phi^2 \rm CS}$ in \cref{eq:EFT_L6}.  
These appear as the leading order contact terms in the EFT $\cM[1_h^- 2_h^- 3_\phi 4_\phi]$ amplitude, as shown  in \cref{eq:EFT_Mhhphiphi}.
We need to calculate the same amplitude at one loop within the scalar matter UV completion and expand it for $E/m_\Phi \ll 1$. 
Its $s$- and $t$-channel discontinuities are related to the tree-level 4-point amplitudes in 
\cref{eq:UV_MhhPhiPhi,eq:UV_MhphiPhiPhi,eq:UV_MphiphiPhiPhi} by
\begin{align}
D_s = &
i \int d \Pi_{XY} \cM[1_h^- 2_h^- (-Y)_\Phi (-X)_{\bar \Phi}] \cM[3_\phi 4_\phi X_{\Phi} Y_{\bar \Phi} ] 
, \nnl 
D_t = &
i \int d \Pi_{XY} \cM[1_h^- 3_\phi (-Y)_\Phi (-X)_{\bar \Phi}] \cM[2_h^- 4_\phi X_{\Phi} Y_{\bar \Phi} ] 
, \end{align} 
while $D_u$ is obtained from $D_t$ by $1 \leftrightarrow 2$. 
By following the same techniques as described above we integrate over the intermediate 2-body phase space and reconstruct the amplitude as   
\begin{align}
\cM[1_h^- 2_h^-3_\phi 4_\phi ]   = & 
\langle 1 2 \rangle^4 
\bigg \{ 
 {m_\Phi^4 \over \mpl^4 s^4 } \bigg [  
 (t^2 - 4 t u + u^2 )  I_{\circ}^s 
-2 t u (2 m_\Phi^2 +s) I_{\triangle}^s    \bigg ]
+ c_\Phi 
{ 2  m_\Phi^4 \over \mpl^2 s^2}   I_{\triangle}^s  
\nnl & 
+ {  y^2  m_\Phi^2  \over \mpl^2 s^2  }  
 \bigg [ 
2 m_\Phi^4 \big [  I_\Box^{st} +  I_\Box^{su} +  I_\Box^{tu} \big ]     
+   t u {4 m_\Phi^2 s + t u \over s^2}   I_\Box^{tu}   
\nnl  &  
- {2  (2 m_\Phi^2 s + t u ) \over s^2}  \big [ t   I_\triangle^t +  u   I_\triangle^u \big ] 
 - { t - u \over s}  
 \bigg ( I_\circ^t
 - I_\circ^u  \bigg )  
  + {\cal R} 
 \bigg ]   
+ c_1 I_1 +{\cal R}
\bigg \} 
. \end{align}  
We expand this for large $m_\Phi$. 
As before, the terms with non-positive powers of 
$E/m_\Phi$ are either non-local and canceled by adjusting $c_1$ and ${\cal R}$, or match to the EFT contributions proportional to the dimension-3 Wilson coefficient $C_3$ and are subject to the rational terms ambiguity. 
We directly focus on the terms with positive powers of 
$E/m_\Phi$ in this expansion. 
We find 
\begin{align}
\label{eq:MATCH_Mhhphiphi-expanded}
\cM[1_h^- 2_h^-3_\phi 4_\phi ]   = & 
\langle 1 2 \rangle^4  
\bigg \{ 
{\color{blue} 
{ t u  \over 40320  \pi^2 \mpl^4 m_\Phi^2 s } 
} 
+ {\color{red} 
{y_0^2 - c_\Phi \over 1440 \pi^2  \mpl^2 m_\Phi^2} 
}
+ {\color{brown}  
{s  \over 6720 \pi^2 \mpl^4 m_\Phi^2  }
}
+ \cO ( m_\Phi^{-4} )
\bigg \} 
. \end{align}
The first term in the curly bracket has a simple pole in the $s$ channel.
Comparing it with the EFT expression in \cref{eq:EFT_Mhhphiphi},
one can read of the Wilson coefficients of the ${\cal R}^3$ dimension-6 operators: 
\begin{align}  
C_{R^3}     = & 
-  { 1  \over 40320  \pi^2  m_\Phi^2 }
, \nnl 
\widetilde C_{R^3}     = & 
0 
,  \end{align}  in agreement with the results of Refs.~\cite{Bern:2021ppb,Alviani:2024sxx} calculated by matching the 4-graviton amplitude.

Furthermore, the second term in the curly bracket of \cref{eq:MATCH_Mhhphiphi-expanded}, which we highlighted in red, is a contact term matching to the 
$C_{\phi^2 \rm GB}$ Wilson coefficient in \cref{eq:EFT_L6}. 
We find 
\begin{align} 
\label{eq:MATCH_cdsgb-scalar}
C_{\phi^2 \rm GB}   =  & 
{y_0^2 - c_\Phi 
\over 1440 \pi^2 m_\Phi^2}  
, \nnl  
C_{\phi^2 \rm CS}   =  & 0
. \end{align} 
Clearly,
$C_{\phi^2 \rm GB}$ is sensitive to the contact term in the UV theory amplitude 
$\cM[1_\phi 2_\phi 3_{\Phi} 4_{\bar \Phi} ]$. 
We can therefore adjust it to any value, positive or negative, unless there is a reason for the $\phi^2 |\Phi|^2$ coupling in the UV Lagrangian to be suppressed compared to $y_0^2$. 
It would however require fine-tuning to set this Wilson coefficient to zero for this particular UV completion. 

Finally, the last term in 
\cref{eq:MATCH_Mhhphiphi-expanded}, marked in brown color, maps to a dimension-8 contact interaction in the EFT Lagrangian. 
This contribution to the corresponding Wilson coefficient $C$ is suppressed by a small factor $m_\Phi/\mpl^2$, and is subleading to another contribution $C \sim y_0^2/m_\Phi^4$ appearing at the next order in the expansion in $E/m_\Phi$.  

In order to determine the dimension-7 $C_{\phi R^3}$ Wilson coefficient we need to match the UV  
$\cM[1_h^- 2_h^- 3_h^- 4_\phi]$ amplitude to the EFT one, 
while $C_4'$ at dimension-8 is read off from the 4-scalar amplitude. 
The calculations are analogous as above and we do not present them explicitly. 
Instead we move directly to the results.

\subsection{Results}
\label{sec:MATCH-results}

Below we match the selected EFT Wilson coefficients of \cref{sec:eft} to the parameters of the UV theory introduced in \cref{sec:uv}. 
The UV parameters are the matter particles masses $m_\Phi$, $m_\Psi$, and $m_V$, as well as a set of the matter couplings $y_i$ to the massless scalar $\phi$. 
We also display the dependence on the leading contact terms in the 4-point amplitude with two massless scalar and two heavy matter fields, which are parametrized by $c_i$ defined in~\cref{eq:UV_MphiphiPhiPhi,eq:AMP_MphiphiPsiPsi,eq:AMP_MphiphiVV}, see the discussion below \cref{eq:UV_MphiphiPhiPhi}. 
For the spin-1/2 and spin-1 matter these contact terms correspond to non-renormalizable operators in the UV theory and are controlled by the cutoff scale $M_*$. 
Other contact terms come with higher powers of $M_*$ and are not relevant for us. 
That same scale enters also through the spin-1 matter couplings to $\phi$, which correspond to non-renormalizable operators as well, cf.~\cref{eq:UV_LUV}.
Finally, the UV Lagrangian may contain the same higher-dimensional operators as in \cref{eq:EFT_Left} but suppressed by suitable powers of the scale $M_*$. 
Therefore all the expressions for the Wilson coefficients below hold up  to $\cO(M_*^{4-D})$, 
where $D$ is the dimension of the corresponding EFT operator. 
The displayed results assume a single particle for each spin, but the generalization is a trivial sum over multiple species.

For the dimension five Wilson coefficients in \cref{eq:EFT_L5} we find 
\begin{align} 
\label{eq:MATCH_csgb}
C_{\phi \rm GB} = & 
-  {y_0  \over 1440  \pi^2   m_\Phi}  
- {7 y_s  \over  2880 \pi^2 m_\Psi}  
- {11 y_d    - 2  y_m \over 240 \pi^2 M_*} 
+ \cO(M_*^{-1})
, \nnl 
C_{\phi \rm CS}  = & 
- { y_p   \over 192 \pi^2 m_\Psi} 
- {y_a  \over 24 \pi^2 M_*} 
+ \cO(M_*^{-1})
.    \end{align} 

For the dimension six Wilson coefficients in \cref{eq:EFT_L6} we find 
\begin{align} 
\label{eq:MATCH_cdsgb}
C_{\phi^2 \rm GB}   =  & 
{y_0^2 - c_\Phi 
\over 1440 \pi^2 m_\Phi^2} 
+ 
{7 (y_s^2  -  y_p^2) \over 2880 \pi^2  m_\Psi^2}   
  + 
{7  c_\Psi \over 2880 \pi^2  M_* m_\Psi }  
\nnl + & 
   { 1 \over  M_*^2 \pi^2}  \bigg \{ 
{5 y_d^2 \over 96} - {y_m^2 \over 768} - {19 y_a^2 \over 480} 
+ {11  c_V \over 240} 
- {11 \hat c_V  \over 1920} 
\bigg \} 
+ \cO(M_*^{-2})
, \nnl  
C_{\phi^2 \rm CS}   =  & 
{y_s y_p \over 96  \pi^2  m_\Psi^2}  
+
{ \widetilde c_\Psi \over 192  \pi^2 M_*  m_\Psi}  
+{ 2 y_d  y_a + \widetilde c_V  
\over  24 M_*^2 \pi^2}
+ \cO(M_*^{-2})
. \end{align} 

Moreover, as a byproduct we recover the results of Refs.~\cite{Goon:2016mil,Bern:2021ppb,Alviani:2024sxx} regarding the matching of the ${\cal R}^3$ operator: 
\begin{align}  
C_{R^3}     = & 
-  { 1  \over 40320  \pi^2  m_\Phi^2 }
+  { 1  \over 20160  \pi^2  m_\Psi^2 } 
-  { 1  \over 13440  \pi^2  m_V^2 } 
+ \cO(M_*^{-2})
, \nnl 
\widetilde C_{R^3}     = & 
 \cO(M_*^{-2}) 
. \end{align} 
The absence of calculable contributions to 
$\widetilde C_{R^3}$ is the consequence of our assumption of minimal gravitational coupling of matter; otherwise, 
$\widetilde C_{R^3}$ could be generated in the presence of the anomalous quadrupole coupling of $V$ to the graviton.  

At dimension seven  we quote the matching for the Wilson coefficients of the operators displayed in \cref{eq:EFT_L7}: 
\begin{align} 
\label{eq:MATCH_csr3}
 C_{\phi \rm R^3}  =  &
{y_0 \over 40320  \pi^2 m_\Phi^3 } 
 - {y_s \over 10080  \pi^2 m_\Psi^3 }
 + {2 y_d + y_m  \over 13440 \pi^2 M_* m_V^2} 
 + \cO(M_*^{-3}) 
, \nnl 
\widetilde C_{\phi \rm R^3} =  & 
\cO(M_*^{-2}) 
. \end{align}  
At dimension eight we quote the matching for the Wilson coefficient of the operator displayed in \cref{eq:EFT_L8}:
\begin{align} 
\label{eq:MATCH_c4prime}
C_4'   =  &
{3 y_0^4 - 8 c_\Phi  y_0^2 + 6 c_\Phi^2 \over 5760 \pi^2 m_\Phi^4}
\nnl + & 
 { 11 y_s^4   
 +38 y_p^2 y_s^2+3 y_p^4 
 \over 1440 \pi^2 m_\Psi^4} 
 + { 11 c_\Psi y_s^2 
-3c_\Psi y_p^2
+30 \widetilde c_\Psi y_s y_p 
\over 720 \pi^2 M_* m_\Psi^3}  
+ { 3 c_\Psi^2 
+5 \widetilde c_\Psi^2 
\over 240 \pi^2 M_*^2 m_\Psi^2} 
   \nnl + &  
\frac{ 32 c_V^2 + 4 \hat{c}_V^2 + 32 \tilde{c}_v^2 + 32 y_a^4 + 64 y_a^2 y_d^2 + 32 y_d^4 - 4 \hat{c}_V y_m^2 + y_m^4
}{
256\pi^2  M_*^4   
} \log\bigg(\frac{\mu^2}{m_V^2}\bigg)
 \nnl + &
 \cO(\mpl^{-2} m^{-2})
+ \cO(M_*^{-4})
. \end{align} 
The vector contribution is UV divergent, therefore 
it is regularization-dependent and depends on the matching scale $\mu$. 
For this reason we only display the log-enhanced part of the result, which is also the one that is free of ambiguities arising from rational terms.

To match the $C_5$ and $C_6$  Wilson coefficients in \cref{eq:EFT_L5,eq:EFT_L6} we would need to calculate the five and six-scalar amplitudes in the UV theory, which we do not attempt in this paper.  

\subsection{Discussion} 

We have seen that heavy matter coupled via cubic interactions to the massless scalar 
offers a framework to UV-complete the SGB  and DCS theories, which are central elements of an important class of  scalar-tensor extensions of GR. 
Below we highlight qualitative features of the matching results displayed in~\cref{sec:MATCH-results}.

A heavy spin-0 particle can provide a simple UV completion for the SGB couplings. 
However, in this setting 
the shift symmetry 
$\phi \to \phi + \alpha$  cannot be implemented.   
In particular, \cref{eq:MATCH_csr3} shows that the shift-symmetry-breaking 
$\phi {\cal R}^3$ operator is always  generated together with $C_{\phi \rm GB}$. 
Moreover, $\phi^2 R^2$ also generically appears, cf.~\cref{eq:MATCH_csr3}, unless it is fine-tuned away by the contact term $c_\Phi$. 
Both positive and  negative values of $C_{\phi^2 \rm GB}$ can be generated, depending on the relative magnitude of the cubic coupling $y_0$ and the contact term. 
We also note that, for $c_{\Phi}= 0$,  the relative value $C_{\phi \rm GB}$ and $C_{\phi^2 \rm GB}$ is consistent with the exponentiation of the massless scalar effective coupling to the Gauss-Bonnet invariant\footnote{%
This often arises from string theory models where the scalar is the dilaton and such a theory is often referred to as Einstein-dilaton-Gauss-Bonnet.}: 
${\cal L}_{\rm EFT} 
\supset
{1 \over 1440 \pi^2} \exp \bigg ( 
- {y_0 \phi \over m_\Phi} 
\bigg ) {\cal R}_{\rm GB}^2$. 

More diversity arises for heavy fermionic matter, which has two possible coupling to the massless scalar: parity-even and shift-symmetry-breaking $y_s$, and  the parity-odd and shift-symmetry-preserving $y_p$. 
If only the former is present, the situation is similar to the one with spin-0 matter. 
In particular, the 
$\phi {\cal R}_{\rm GB}^2$ coupling is generated but $\phi {\cal R}_{\rm CS}^2$ is not, and shift symmetry is violated by various EFT operators. 
Note that once again the relative values of $C_{\phi \rm GB}$ and $C_{\phi^2 \rm GB}$ are consistent with  exponentiation, as in 
${\cal L}_{\rm EFT} 
\supset
{7 \over 2880 \pi^2} \exp \bigg ( 
- {y_s \phi \over m_\Psi} 
\bigg ) {\cal R}_{\rm GB}^2$.

For $y_s =0$ and $y_p \neq 0$ the pattern of the EFT coefficients is qualitatively different. 
As expected on the symmetry grounds, in this case it is the 
$\phi {\cal R}_{\rm CS}^2$ coupling that is generated, while the 
$\phi {\cal R}_{\rm GB}^2$ one is not. 
Moreover, the EFT now can inherit the shift symmetry of the parent theory.
Indeed, \cref{eq:MATCH_csr3} shows that the $\phi {\cal R}^3$ operator is not generated through $y_p$. 
The vanishing of 
$C_{\phi^2 \rm GB}$ in this limit is not automatic, as it requires tuning the contact term as 
$c_\Psi = {M_* \over m}$, putting the cutoff of the UV theory close to the scale $m$. 
But precisely this contact term is required if $\phi$ is interpreted as a Goldstone boson of a spontaneously broken Peccei-Quinn global symmetry $U(1)_{\rm PQ}$.  
Starting from a model with a complex scalar field $H$  and  the interactions 
  \begin{align}  
  \label{eq:MATCH_LU1PQ}
{\cal L} \supset & 
-  \sqrt 2  y_p \big [  
H   \bar{\Psi}_R \Psi_L  +  \hc  \big ] 
 ,  \end{align}  
the action of  
$U(1)_{\rm PQ}$ on the fields can be  defined as 
\begin{align}   
 H \to e^{i \alpha } H
 , \qquad 
\Psi_L \to e^{- i \alpha } \Psi_L  
 , \qquad 
 \Psi_R \to \Psi_R 
.   \end{align}  
In particular, 
$U(1)_{\rm PQ}$ acts chirally on the heavy  fermion $\Psi$. 
Once the scalar gets a vacuum expectation value,  $\langle H \rangle = f$,  
 $U(1)_{\rm PQ}$ is spontaneously broken, giving rise to one massless Goldstone boson. 
To isolate it we write 
$H  = {f + h \over \sqrt 2} e^{i \phi/f}$ so that  
$U(1)_{\rm PQ}$ acts as the shift $\phi \to \phi + \alpha$. 
Then, integrating out $h$ leaves 
 \begin{align}  
{\cal L} \supset & 
-  y_p  f    \big [   e^{- i \phi/f}  \bar \Psi_R \Psi_L  +  e^{ i \phi/f}  \bar \Psi_L \Psi_R \big ] 
\nnl = & 
-  m_\Psi (\bar \Psi \Psi) 
+ i y_p  \phi  (\bar \Psi \gamma_5 \Psi) 
+  {y_p^2 \over m_\Psi}   {\phi^2 \over 2} (\bar \Psi \Psi) 
+ \cO(\phi^3)
 ,  \end{align}   
where $m_\Psi = y_p f$. 
The quadratic term in $ \phi$ manifests as the contact terms in the amplitude of \cref{eq:AMP_MphiphiPsiPsi} with 
$c_\Psi = 
{M_* \over m_\Psi}$ 
and 
$\widetilde c_\Psi  = 0$, as required by the vanishing of the shift-symmetry-violating coefficients $C_{\phi^2 \rm GB}$ and $C_{\phi^2 \rm CS}$ in the EFT. 

The model discussed above was considered from a different perspective in Ref.~\cite{Alexander:2022cow}. 
As discussed in that reference,  the precise magnitude of 
$C_{\phi \rm CS}$ in \cref{eq:MATCH_csgb} can be understood by anomaly matching between the EFT and the UV theory. 
In the UV theory described by \cref{eq:MATCH_LU1PQ}, 
$U(1)_{\rm PQ}$ has a mixed anomaly with gravity, 
such that the 
$U(1)_{\rm PQ}$ symmetry current satisfies 
$\partial_\mu J^\mu = 
{1 \over 384 \pi^2} 
R_{\mu \nu \alpha \beta} 
\widetilde R^{\mu \nu \alpha \beta}$. 
The same right-hand side is reproduced in the EFT, taking into account the DCS interaction in \cref{eq:EFT_L5}, the matching in \cref{eq:MATCH_csgb}, and the $\phi \to \phi + \alpha$  transformation of the massless scalar under $U(1)_{\rm PQ}$. 
We refer the reader also to~Refs.~\cite{Bonnefoy:2020gyh,Alexander:2024vav} for related discussions on the emergence of axion-like couplings such as the DCS one. 

The case of UV completion with vector matter is again qualitatively different. 
In this case the matter coupling to the massless scalar proceeds via non-renormalizable operators suppressed by $M_* \gg m_V$, which sets the scale of new particles that UV-complete the UV theory of \cref{sec:uv}. 
The contributions to the SGB and DCS couplings from integrating out spin-1 matter are then suppressed by this high scale, cf.~\cref{eq:MATCH_csgb,eq:MATCH_cdsgb}, rather than by the matter particle mass as for the spin-0 and spin-1/2 matter. 
These  are of the same order as the unknown contributions from integrating out particles at the scale $M_*$. 
At the same time, for our method,  uncontrolled contributions from rational terms are of the same order,  as only effects scaling with negative powers of matter masses are not subject to the rational terms ambiguity.  
The conclusion is that the UV completion with spin-1 matter, while generically expected to produce the SGB and DCS couplings in the EFT, is not predictive. 
We expect this feature will persist for UV theory with matter of spin higher than one.

Our discussion reveals the difficulty of generating solely the $\phi {\cal R}_{\rm GB}^2$ coupling at one loop from a shift-invariant UV completion. 
For all the UV setups considered, either the 
$\phi {\cal R}_{\rm GB}^2$ coupling is absent in the EFT (when only the parity-odd couplings $y_p$ and $y_a$ are present in the UV) or shift-symmetry breaking operators are generated along with $C_{\phi \rm GB}$. 
Generically, lower-dimensional operators such as $\phi^3$ and $\phi^4$ will also be present in the EFT, unless they are fine-tuned away. 
On the other hand, the 
$\phi {\cal R}^2_{\rm CS}$ coupling can readily  be generated from a shift-invariant UV completion, with a massless scalar identified as a Goldstone boson of $U(1)_{\rm PQ}$ under which heavy fermionic matter has chiral transformations.  

Overall, for the spin-0 and spin-1/2 UV completion the EFT fits the following form:
\begin{align}
{\cal L}_{\rm EFT} = & 
{\mpl^2  R  \over 2} 
+  {(\partial_\mu \phi)^2  \over 2} 
+{\Lambda^4 \over 
16 \pi^2  } 
L \bigg [ {\partial_\mu \over \Lambda }, 
{R_{\mu \nu \alpha \beta} \over \Lambda^2}, 
{g \phi \over \Lambda} 
\bigg ] 
, \end{align}
where $\Lambda$ is the EFT cutoff given by the scale of matter particle masses, 
and $g$ is of the order of the cubic couplings in the UV theory. 
This is in agreement with power counting rules of Ref.~\cite{Serra:2022pzl} for one-loop generated operators. 
For spin-1 UV completion one should replace 
${g \phi \over \Lambda} \to {g \phi \over M_*}$.

A closing remark concerns the relationship between the EFT Wilson coefficients we obtained and the constraints coming from locality, causality, and Poincar\'{e} invariance of the underlying theory~\cite{Adams:2006sv}. 
The UV theory we consider obviously respects these requirements, therefore the Wilson coefficients of our EFT must satisfy the corresponding family of  constraints. 
One example is the positivity constraint on the dimension-8 coefficient of the 4-scalar interactions 
$C_4'> 0$~\cite{Adams:2006sv}, 
which can be easily shown to be satisfied, up to $\cO(\mpl^{-2} m^{-2})$ effects, by \cref{eq:MATCH_c4prime} in the entire parameter space.
Furthermore, Ref.~\cite{Serra:2022pzl} considered causality bounds on the massless scalar couplings to gravity that follow from requiring positive Shapiro time delay within the EFT validity regime~\cite{Camanho:2014apa}\footnote{More discussion on causality and positivity bounds for these theories can also be found in~Refs.~\cite{Hong:2023zgm,Xu:2024iao,Nie:2024pby}}. 
The conclusion is that the shift-symmetric SGB and DCS Wilson coefficients are bounded as 
\begin{align}
\label{eq:MATCH_causalitybound}
\hat C \equiv 
|C_{\phi \rm GB} 
+ i C_{\phi \rm CS}| \lesssim {\mpl \over \Lambda^2}
. \end{align}    
Assuming sub-Planckian matter masses,  
this bound is trivially satisfied by \cref{eq:MATCH_csgb} for a single matter particle. 
On the other hand, for a large number $N$ of matter species, assuming maximum cubic coupling consistent with perturbative unitarity of the UV theory, 
$N y^2 \sim 1$,  ignoring numerical and $\pi$ factors 
one has 
$\hat C \sim {N y \over \Lambda} \sim {\sqrt{N} \over \Lambda}$. 
This is consistent with 
\cref{eq:MATCH_causalitybound} provided $N \lesssim 
{\mpl^2 \over \Lambda^2}$.

\section{Conclusions} 
\label{sec:con}

In this paper we studied a UV completion to  scalar-tensor EFTs of gravity featuring a massless spin-0 particle $\phi$ in addition to the spin-2 graviton. 
We focused on the scalar having a cubic coupling with two gravitons, described in the Lagrangian language by dimension-5 operators coupling  $\phi$ to the Gauss-Bonnet and Chern-Simons invariants. 
We embedded this EFT into a more fundamental theory where the SGB and DCS couplings are absent, and which features massive matter fields  with spin $S=0,1/2,1$.   
Integrating out the heavy matter at one loop generates the leading SGB and DCS interactions in the effective theory below the matter mass scale.
Our main results, summarized in \cref{sec:MATCH-results}, are the Wilson coefficients of the corresponding dimension-5 EFT operators expressed in terms of the masses and couplings in the UV theory. 
We also calculated the Wilson coefficients of a number  dimension-6 and -7 operators that are generically generated along with the $\phi {\cal R}_{\rm GB}^2$ and $\phi {\cal R}_{\rm CS}^2$ couplings.   

Our main focus was on the  pattern of Wilson coefficients in the effective theory below the heavy matter mass scale, in particular on how symmetries of the UV theory are reflected in the EFT.
For example, the shift-symmetric version of the DCS theory can be considered an EFT UV-completed by a massive spin-1/2 fermion $\Psi$ with the pseudoscalar coupling to the massless scalar. 
The UV theory possess a shift symmetry acting as $\phi \to \phi + \alpha$, which can be inherited by EFT provided $\phi$ is a Goldstone boson of a spontaneously broken Peccei-Quinn symmetry acting chirally on $\Psi$. 
In such a case, the dimension-5 operator responsible for the DCS coupling, 
$C_{\phi \rm CS} \sim {1 \over (4 \pi)^2 m_\Psi}$,  is the lowest-dimensional deformation of the EFT after the Einstein-Hilbert term (dimension two) and the scalar kinetic term (dimension four). Further higher-dimensional operators arise at dimension seven and above. In particular, quartic self-interactions of the massless scalar appear from a derivative operator of dimension eight, with a positive Wilson coefficient up to corrections suppressed by the Planck scale. 

On the other hand, the shift-symmetric SGB theory does not have shift-symmetric UV completions within the considered class of models. 
Integrating out heavy matter at one loop does produce the corresponding dimension-5 operator 
$\phi {\cal R}_{\rm GB}^2$ with 
$C_{\phi \rm GB} \sim 
{1 \over (4 \pi)^2 m_{\rm matter}}$, but generically it also produces a host of other operators at lower or higher dimensions, e.g. $\phi^n$, $\phi^2 {\cal R}_{\rm GB}^2$ , $\phi {\cal R}^3$, etc. 
While these extra operators can be fine-tuned away (using contact terms or cancellation between contributions from different particles), within the considered class of UV completions there is no symmetry principle protecting the structure of SGB theories considered in the literature.

The results we have presented shed new light on the possible fundamental origins of scalar-tensor theories of gravity. 
They also stress the EFT nature of such theories. 
The consequence is that the shift-symmetric SGB and DCS interactions are generically accompanied by other interactions, whose strength can be deduced by EFT power counting. 
It would be interesting to perform a systematic study of black hole solutions and scalar hair in the presence of the additional operators predicted by known UV completions. 
Exploring the effects of the operators $\phi^2 {\cal R}^2_{\rm GB}$ is especially relevant, given their role in triggering spontaneous scalarization~\cite{Doneva:2017bvd,Silva:2017uqg}.
Another interesting extension of our study would be to check how the inclusion of non-minimal couplings of the matter particles to gravity can alter the conclusions of this paper. 
Lastly, a potentially fruitful direction would be to investigate whether the Wilson coefficients of our low-energy EFT can also be obtained from string theory amplitudes — similarly to what was done in Ref.~\cite{Bern:2021ppb} for the ${\cal R}^3$ and ${\cal R}^4$ operators — and whether any structure or pattern emerges in how they relate to the coefficients obtained in this work.

\appendix 

\section{Phase space integration} 
\label{app:PSI}

In this appendix we summarize the techniques for dealing with 2-body phase space integrals appearing in the calculation of one-loop scattering of massless gravitons and scalars with heavy matter particles in the loop.

\subsection{Parametrization}

For $s$-channel cuts, the massive internal momenta $p_X$ and $p_Y$ 
can be parametrized as
\begin{align}
\label{eq:PSI_pXYofalphaz}
p_X= & 
\alpha p_1+(1-\alpha) p_2-\sqrt{\alpha(1-\alpha)-\frac{m^2}{s}}[zq+z^{-1} \bar q]
    ,\notag\\
p_Y= &
(1-\alpha) p_1+\alpha p_2+\sqrt{\alpha(1-\alpha)-\frac{m^2}{s}}[zq+z^{-1} \bar q]
, \end{align}
where $p_1$ and $p_2$ are  massless momenta, $p_1^2=p_2^2=0$, 
and $q \sigma = |2\rangle [1|$, 
$\bar q \sigma = |1\rangle [2|$. 
Given these definitions, the mass-shell conditions  $p_X^2 = p_Y^2 = m^2$ follow automatically. 
The parameter $\alpha$ is in the range 
\begin{equation}
\alpha\in[\alpha_-,\alpha_+],\qquad\alpha_\pm=\frac{1\pm\sqrt{1-\frac{4m^2}{s}}}{2},
\end{equation}
and the other parameter is constrained to the unit circle, $|z|=1$. 
The 2-body phase space element for the $X$ and $Y$ pair is given by 
\begin{align}
\text{d}\Pi_{XY}=
\frac{\text{d}\alpha}{8\pi}
\frac{\text{d}z}{2\pi i z}
.\end{align}
In the limit $m \to 0$ the parametrization in \cref{eq:PSI_pXYofalphaz} reduces to the one 
introduced in Refs.~\cite{Zwiebel:2011bx,Caron-Huot:2016cwu} with $\alpha =\cos^2 \theta$, $z=e^{i\phi}$.

\subsection{Scalar integrals}

The one-loop amplitude we calculate are expressed in the basis of scalar integrals defined as 
\begin{align}
\label{eq:PSI_scalar-integrals}
I_1 \equiv  & 
\int\frac{\text{d}^4 k}{i(2\pi)^4}\frac{1}{[k^2-m^2]}
,\notag\\
I_\circ^s \equiv  & \int\frac{\text{d}^4 k}{i(2\pi)^4}\frac{1}{[k^2-m^2][(k+p_1+p_2)^2-m^2]}
,\notag\\
I_\triangle^s \equiv  & 
\int\frac{\text{d}^4 k}{i(2\pi)^4}\frac{1}{[k^2-m^2][(k+p_1+p_2)^2-m^2][(k+p_1)^2-m^2]}
,\notag\\  
I_\Box^{st} \equiv  & 
\int\frac{\text{d}^4 k}{i(2\pi)^4}\frac{1}{[k^2-m^2][(k+p_1)^2-m^2][(k+p_1+p_2)^2-m^2][(k-p_3)^2-m^2]}
. \end{align}
The tadpoles and bubbles are implicitly dimensionally regularized with $d=4-2\epsilon$. 
In addition, the basis includes 
$I_\circ^t$, $I_\circ^u$, 
$I_\triangle^t$, $I_\triangle^u$,
$I_\Box^{su}$, 
and $I_\Box^{tu}$,
which can be obtained from the above definitions by crossing 
($s \leftrightarrow t$ and 
$2 \leftrightarrow 3$, 
or 
$s \leftrightarrow u$ and 
$2 \leftrightarrow 4$). 

The superscript in \cref{eq:PSI_scalar-integrals} indicates the  channels in which the integral have discontinuities. 
Using the parametrization in \cref{eq:PSI_pXYofalphaz} these can be concisely expressed as integrals over $\alpha$ parameters: 
 \begin{align}         
\label{eq:PSI_alphaIntegrals-s}  
 {\rm Disc}_s    I_\circ^s      = &      
    { i  \over 8 \pi}    \int_{\alpha_-}^{\alpha_+}  d\alpha
, \nnl 
{\rm Disc}_s I_\triangle^s  = &  
- {i \over 8 \pi }  \int_{\alpha_-}^{\alpha_+} {d\alpha    \over  \sqrt{ [u \alpha   -t  (1- \alpha) ]^2  + 4 m^2 { t u \over  s }  }  }
, \nnl  
 {\rm Disc}_s   I_\Box^{st}  =  & 
  { i\over 8 \pi  s   }   \int_{\alpha_-}^{\alpha_+}  
 { d\alpha  \over \alpha \sqrt{ [u \alpha   - t  (1- \alpha) ]^2  + 4 m^2  { t u  \over  s }  } } 
 , \nnl  
 {\rm Disc}_s   I_\Box^{su}  =  & 
  { i\over 8 \pi  s   }   \int_{\alpha_-}^{\alpha_+}  
 { d\alpha  \over (1- \alpha) \sqrt{ [u \alpha   - t  (1- \alpha) ]^2  + 4 m^2  { t u  \over  s }  } } 
 .  \end{align}  
The integrals can be easily performed explicitly, expressing the discontinuities in terms of logarithms and square roots of Mandelstam invariants polynomials. 
In practice, however, it is easier to work at the integrand level. 
For example, starting with the expression in \cref{eq:MATCH_DsMhphiphiphi}, trading  the $p_X$ variable for $\alpha$ and $z$, integrating over $z$, one can recast the integrand as a sum of those in \cref{eq:PSI_alphaIntegrals-s}
to arrive at \cref{eq:MATCH_DsMhphiphiphi-result}.

\section{Relevant amplitudes} 
\label{sec:amp}

\subsection{Tree level}

We collect here the tree-level helicity amplitudes in the UV theory describing scattering of massless scalars and gravitons into  a 
pair of heavy matter particles.  
These serve as building blocks to construct one loop amplitudes for the scattering of massless degrees of freedom in the UV theory. 
The relevant amplitudes with spin-0 heavy matter were given in \cref{eq:UV_MhhPhiPhi,eq:UV_MhphiPhiPhi,eq:UV_MphiphiPhiPhi}. 
Below we give the analogous expressions for spin-1/2 and spin-1 heavy matter pairs:

{\bf Spin 1/2.} 

 \begin{align}
\cM[1_h^- 2_h^+ 3_\Psi 4_{\bar \Psi} ]  =  &
{\langle1|p_3|2]^3 
\big ( \langle 1 \bm 3\rangle [2 \bm 4] + \langle 1 \bm 4\rangle [2 \bm 3]  \big ) 
\over \mpl^2 s (t-m_\Psi^2)(u-m_\Psi^2) } 
, \nnl 
\cM[1_h^- 2_h^- 3_\Psi 4_{\bar \Psi} ]  =  &
{m_\Psi^3 \langle 1 2 \rangle^4 [\bm3 \bm 4] \over \mpl^2 s (t-m_\Psi^2)(u-m_\Psi^2) } 
. \end{align}  
 \begin{align}
\cM[1_h^- 2_\phi 3_\Psi 4_{\bar \Psi}  ]  =  &
 {(y_s + i y_p) \langle 1| p_2 p_3|1\rangle  \over \mpl (t-m_\Psi^2)(u-m_\Psi^2)} \bigg \{ 
{ \langle 1| p_2 p_3|1\rangle  \over s  }
\big ( \langle \bm 3 \bm 4\rangle + \eta [ \bm 3 \bm 4] \big )
-  \langle 1 \bm 3 \rangle \langle 1 \bm 4 \rangle  
 \bigg \} 
, \end{align}
\begin{align}   
\label{eq:AMP_MphiphiPsiPsi}
\cM[1_\phi 2_\phi 3_\Psi 4_{\bar \Psi} ] =  & 
  { t - u \over 2   \mpl^2 s  }   \big ( [\bm 3|p_1| \bm4\rangle+\langle \bm 3|p_1|\bm 4] \big )  
-{ (y_s^2 + y_p^2) (t  - u)   \over (t -  m_\Psi^2)(u -  m_\Psi^2)  }    
     \big (  \tilde [\bm 3|p_1|\bm4\rangle + \langle \bm 3|p_1| \bm 4]  \big )
\nnl  +  &
{ 2 y_s^2 m_\Psi s \over (t -  m_\Psi ^2)(u -  m_\Psi^2)  }    \big ( [ \bm 3 \bm 4] + \langle\bm 3 \bm 4 \rangle \big )
+  {2 i y_s y_p m_\Psi s \over (t -  m_\Psi ^2)(u -  m_\Psi^2)  } \big ( \langle \bm 3 \bm 4 \rangle - [\bm3 \bm4] \big )
 \nnl  +  & 
 {c_\Psi \over M_*}   \big ( \langle \bm 3 \bm 4 \rangle + [\bm 3 \bm 4]  \big )
+   {i \tilde c_\Psi  \over M_*}   \big ( \langle \bm 3 \bm 4 \rangle - [ \bm 3 \bm 4] \big )  
.  \end{align} 

{\bf  Spin-1}. 
\begin{align}
\cM[1_h^- 2_h^+ 3_V 4_{\bar V} ]  =  &
{ \langle 1|p_3|2]^2  
\big ( \langle 1 \bm 3\rangle [2 \bm 4] + \langle 1 \bm 4\rangle [2 \bm 3] \big )^2 
\over \mpl^2 s (t-m^2)(u-m^2) } 
, \nnl 
\cM[1_h^- 2_h^- 3_V 4_{\bar V} ]  =  &
{m^2 \langle 1 2 \rangle^4 [\bm3 \bm 4]^2  \over \mpl^2 s (t-m^2)(u-m^2) } 
, \end{align}  
\begin{align}
  \cM[1_h^- 2_\phi 3_V 4_{\bar V}  ]  = &
     {1  \over \mpl m (t -m^2) ( u-m^2)  }
\bigg \{
 \nnl  & 
 {   \langle 1 | p_3p_2|1\rangle^2  \over s}  
 \bigg ( 
y_d   \big (  \langle \bm 3 \bm4 \rangle^2 + [\bm 3 \bm 4]^2     \big )  
+ y_m   \langle \bm 3 \bm4 \rangle  [\bm 3 \bm 4]
+ i y_a   \big (  \langle \bm 3 \bm4 \rangle^2 - [\bm 3 \bm 4]^2     \big )    
\bigg ) 
 \nnl & 
 +   \langle 1 | p_3p_2|1\rangle \langle 1 \bm 3 \rangle \langle 1 \bm 4 \rangle   
 \bigg ( 2 (y_d   +  i y_a) \langle \bm 3 \bm4 \rangle + y_m [ \bm 3 \bm4 ]  \bigg ) 
 + (y_d   +  i y_a)   s \langle 1 \bm 3 \rangle^2 \langle 1 \bm 4 \rangle^2 
 \bigg  \}  
. \end{align} 
\begin{align}   
\label{eq:AMP_MphiphiVV}
  \cM[1_\phi 2_\phi 3_V 4_{\bar V} ] =&   
 {t - u \over 2 \mpl^2  s } \bigg \{  
{ t - u \over 2 m }  \langle \bm 3 \bm 4 \rangle [ \bm 3 \bm 4]  
+   \big ( [\bm 3|p_1|\bm 4\rangle + \langle \bm 3|p_1| \bm 4 ]  \big )  
 \big (\langle \bm 3 \bm 4 \rangle  +  [ \bm 3 \bm 4 ] \big )  \bigg \} 
 \nnl + & 
{  s  m^2   \over M_*^2 (t - m^2)(u-m^2)  }  \bigg \{   
2 y_d^2   \big(  \langle \bm 3 \bm 4 \rangle^2  +  [\bm 3 \bm 4]^2  \big ) 
+ y_m^2   \langle \bm 3 \bm 4 \rangle [\bm 3 \bm 4]     
 \nnl &   
 + y_m y_d   \big (   \langle \bm 3 \bm 4 \rangle  +    [\bm 3 \bm 4]     \big )^2 
 +  i y_a (2  y_d + y_m) \big (         \langle \bm 3 \bm 4 \rangle^2  - [\bm 3 \bm 4]^2  \big ) 
 \nnl &  
+  {y_d^2 + y_a^2 \over m^2 }  \big (  \langle \bm3|p_1|\bm4]^2+[\bm3|p_1|\bm4\rangle^2   \big )
+ {y_m^2 \over 2  m^2   } \langle \bm 3|p_1 | \bm 4] [ \bm 3|p_1|\bm 4\rangle     
  \bigg \} 
 \nnl  + & 
 {m \over M_*^2}   \bigg \{  
 2 (y_d^2 + y_a^2)  \bigg  [  { \langle \bm 3 |p_1 | \bm 4] [\bm 3 \bm 4] + [\bm 3|p_1|\bm4 \rangle \langle \bm 3 \bm 4 \rangle    \over t -m^2} 
- { \langle \bm 3|p_1 | \bm 4] \langle \bm 3 \bm 4 \rangle + [\bm 3|p_1| \bm 4 \rangle [ \bm 3 \bm 4]  \over 
 u - m^2 } \bigg ]
  \nnl  &  
  + {y_m^2 \over 2 }  \bigg [    
     { \langle \bm 3 | p_1 | \bm 4] \langle \bm 3 \bm 4 \rangle + [ \bm 3 | p_1 | \bm 4 \rangle [ \bm 3 \bm 4 ]     
\over t - m^2 } 
- {  \langle \bm 3 | p_1 | \bm 4] [ \bm 3 \bm 4] + [ \bm 3 | p_1 | \bm 4 \rangle \langle \bm 3 \bm 4 \rangle   
\over u - m^2 }  \bigg ] 
  \nnl  &-  { y_m (t - u) \over  ( t - m^2)  (u - m^2) }
   \big (  \langle \bm 3 |p_1 | \bm 4] + [ \bm3|p_1| \bm 4 \rangle   \big )  
  \bigg [  y_d   \big ( \langle \bm 3 \bm 4 \rangle + [ \bm 3 \bm 4]  \big ) 
  + i y_a \big ( \langle \bm 3 \bm 4 \rangle - [ \bm3 \bm 4] \big ) \bigg ] 
  \bigg \}   
  \nnl  + &  
{1 \over M_*^2}     \bigg \{ 
 c_V  \big [ \langle \bm 3 \bm 4 \rangle^2  +  [ \bm3 \bm 4]^2 \big ]   
  +   i \tilde c_V     \big [ \langle \bm 3 \bm 4 \rangle^2  -   [ \bm3 \bm 4]^2 \big ] 
  + \hat c_V    \langle \bm 3 \bm 4 \rangle   [ \bm3 \bm 4]   \bigg \} 
.  \end{align}  

\subsection{One loop}

Here we list the UV one-loop amplitudes needed to compute the matching results in \cref{sec:MATCH-results}. The rational terms and tadpoles are not fixed by unitarity, but tadpoles and the $\cO(m_X^{-n})$ parts of the rational terms can be determined requiring that the theory has a smooth $m_X\rightarrow 0$ limit. \\
{\bf  Spin-0}. 
\begin{align}
\label{eq:SUV_Mhhhphi}
\mathcal{M}[1_h^- 2_h^- 3_h^+ 4_\phi] =& 
\frac{\langle 12 \rangle^4[3| p_1  p_2  |3]^2}
     {M_{\text{Pl}}^3 s t u} 
\frac{m_\Phi y_0}{s^2}
\bigg\{ \nonumber \\
& 2 m_\Phi^6 \left[ I_{\Box}^{st} + I_{\Box}^{su} + I_{\Box}^{tu} \right] 
+ \frac{m_\Phi^4 s t}{u} I_{\Box}^{st} + \frac{m_\Phi^4 s u}{t} I_{\Box}^{su} \nonumber \\
& + t u \frac{9 m_\Phi^4 s^2 + 6 m_\Phi^2 s t u + t^2 u^2}{s^3} I_{\Box}^{tu} \nonumber \\
& + 2 \frac{m_\Phi^4 s^2(s^4 - 3 t^3 u) - 4 m_\Phi^2 s t^4 u^2 - t^5 u^3}
         {s^3 t^2 u} I_{\triangle}^t \nonumber \\
& + u \frac{2 m_\Phi^2 s (5 s^2 - 10 s t + 18 t^2) 
      - t (s^3 - 2 s^2 t + 6 s t^2 + 12 t^3)}
       {6 s^2 t^2} I_{\circ}^t \nonumber \\
& + 2 \frac{m_\Phi^4 s^2(s^4 - 3 u^3 t) - 4 m_\Phi^2 s u^4 t^2 - u^5 t^3}
         {s^3 u^2 t} I_{\triangle}^u \nonumber \\
& + t \frac{2 m_\Phi^2 s (5 s^2 - 10 s u + 18 u^2) 
      - u (s^3 - 2 s^2 u + 6 s u^2 + 12 u^3)}
       {6 s^2 u^2} I_{\circ}^u \nonumber \\
& + 2 m_\Phi^4 \left( \frac{s^2}{t u} - 1 \right) I_{\triangle}^s 
+ c_1 I_1 + \mathcal{R}
\bigg\}.
\end{align}

\begin{align} 
\label{eq:SUV_Mhhhphi-allminus} 
\mathcal{M}[1_h^- 2_h^- 3_h^- 4_\phi] =& 
\frac{\langle 12 \rangle^2 \langle 13 \rangle^2 \langle 23 \rangle^2}{M_{\text{Pl}}^3}  
\frac{y_0 \, m_\Phi^5}{s t u} 
\bigg\{ \nonumber \\
& 2 m_\Phi^2 \left[ I_{\Box}^{st} + I_{\Box}^{su} + I_{\Box}^{tu} \right] 
+ \frac{s t}{u} I_{\Box}^{st}
+ \frac{s u}{t} I_{\Box}^{su}
+ \frac{t u}{s} I_{\Box}^{tu} \nonumber \\
& + 2 \frac{t^2 + t u + u^2}{t u} I_{\triangle}^s 
+ 2 \frac{s^2 + s u + u^2}{s u} I_{\triangle}^t 
+ 2 \frac{t^2 + t s + s^2}{t s} I_{\triangle}^u 
+ \mathcal{R}
\bigg\}.
\end{align}

\begin{align}
\label{eq:SUV_Mhhphiphi}
\cM[1_h^- 2_h^- 3_\phi 4_\phi] = & 
\langle 1 2 \rangle^4 
\bigg \{ 
\frac{m_\Phi^4 }{\mpl^4 s^4} 
\bigg [ 
(t^2 - 4 t u + u^2) I_{\circ}^s 
- 2 t u (2 m_\Phi^2 + s) I_{\triangle}^s 
\bigg ] 
+ \frac{2 c_{\Phi} m_\Phi^4 }{\mpl^2 s^2} I_{\triangle}^s 	 \nonumber \\
& + {  y_0^2  m_\Phi^2  \over \mpl^2 s^2  }  
	\bigg [ 
	2 m_\Phi^4 \big [  I_\Box^{st} +  I_\Box^{su} +  I_\Box^{tu} \big ]     
	+   t u {4 m_\Phi^2 s + t u \over s^2}   I_\Box^{tu}   
	\nnl  & 
	- {2  (2m_\Phi^2 s + t u ) \over s^2}  \big [ t   I_\triangle^t +  u   I_\triangle^u \big ] 
	- {s + 2 t \over s}   I_\circ^t
	- {s + 2 u \over s}   I_\circ^u 
	\bigg  ]  + c_1 I_1 +{\cal R} 
    \bigg \} 
\end{align}

\begin{align}
\cM[1_\phi 2_\phi 3_\phi 4_\phi] = & 
	2 y_0^4 m_\Phi^4  \big [  I_\Box^{st} + I_\Box^{su} + I_\Box^{tu} \big ] 
	+ \bigg \{ 4 y_0^2 m_\Phi^2  \bigg [ c_\Phi - { (2 m^2 + s) t u \over \mpl^2 s^2 } \bigg ] I_\triangle^s 
	\nnl & + \bigg [ 
	c_\Phi^2 
	+\frac{6 m_\Phi^2 y^2 \left(t^2-4 t u+u^2\right)  - s^2 c_\Phi \left(2 m_\Phi^2+s\right)}{ 3 \mpl^2 s^2}
	\nnl  & 
	+\frac{2 m_\Phi^2 s \left(t^2+6 t u+u^2\right)+m_\Phi^4 \left(6 t^2-4 t u+6 u^2\right)
		+s^2 \left(t^2+t u+u^2\right)}{30 \mpl^4 s^2}
	\bigg ]I_\circ^s \nnl &
    +(s\leftrightarrow t) + (s\leftrightarrow u) 
	\bigg \} +c_1 I_1 + \mathcal{R}  
	. \end{align}

{\bf  Spin-1/2}. 
\begin{align}
\label{eq:FUV_Mhhhphi}
\mathcal{M}[1_h^- 2_h^- 3_h^+ 4_\phi] =& 
\frac{\langle 12 \rangle^4 [3 |p_1  p_2  |3]^2}
     {M_{\text{Pl}}^3 s t u} 
\frac{m_\Psi}{s^2} 
\bigg\{ \nonumber \\
& m_\Psi^2 
\frac{(2 m_\Psi^2 u + s t) 
      [s (y_s + y_p) - 4 m_\Psi^2 y_s]}{u} 
\, I_{\Box}^{st} \nonumber \\
& + m_\Psi^2 
\frac{(2 m_\Psi^2 t + s u) 
      [s (y_s + y_p) - 4 m_\Psi^2 y_s]}{t} 
\, I_{\Box}^{su} \nonumber \\
& + \bigg[
      \left( 2 m_\Psi^4 s + 4 m_\Psi^2 t u + \frac{t^2 u^2}{s} \right)(y_s + y_p)
      \nonumber \\
& \qquad 
      - 2 y_s \left( 
         4 m_\Psi^6 + \frac{18 m_\Psi^4 t u}{s} 
         + \frac{12 m_\Psi^2 t^2 u^2}{s^2} 
         + \frac{2 t^3 u^3}{s^3} 
      \right)
   \bigg] 
   I_{\Box}^{tu} \nonumber \\
& - \bigg[
      \left( 
         2 m_\Psi^2 \frac{s^3 - 2 s t^2 - 2 t^3}{t u} 
         + \frac{2 t^2 u}{s} 
      \right)(y_s + y_p)
      \nonumber \\
& \qquad 
      + \frac{8 y_s}{s^3 t^2 u} \big(  
         m_\Psi^4 s^2 (s^4 + 3 s t^3 + 3 t^4) 
         - 4 m_\Psi^2 s t^4 u^2 
         - t^5 u^3  
      \big) 
   \bigg] 
   I_{\triangle}^t + \left[ t \leftrightarrow u \right] I_{\triangle}^u \nonumber \\
& + u \bigg[
      \left( 2 - \frac{s}{t} \right)(y_s + y_p)
      \nonumber \\
& \qquad 
      + \frac{2 y_s}{3 s^2 t^2} \big( 
         -2 m_\Psi^2 s (5 s^2 - 10 s t + 18 t^2) 
         + t (s^3 - 2 s^2 t + 6 s t^2 + 12 t^3) 
      \big)
   \bigg] 
   I_{\circ}^t \nonumber \\
& + t \left[ t \leftrightarrow u \right] I_{\circ}^u + 2 m_\Psi^2 
      \left[ s (y_s + y_p) - 4 m_\Psi^2 y_s \right] 
      \left( \frac{s^2}{t u} - 1 \right) 
      I_{\triangle}^s 
+ c_1 I_1 + \mathcal{R} 
\bigg\}.
\end{align}

\begin{align} 
\label{eq:FUV_Mhhhphi-allminus}
\cM[1_h^- 2_h^-3_h^- 4_\phi ]    =& 
- 4
{\langle 12 \rangle^2 \langle 13 \rangle^2   \langle 23 \rangle^2 \over  \mpl^3   }  
{ y_s m_\Psi^5  \over s t u} 
\bigg \{ 
2m_\Psi^2 \big [ I_{\Box}^{st} +  I_{\Box}^{su} +I_{\Box}^{tu} \big ] 
+ {s t  \over u}  I_{\Box}^{st}
+ { s u \over t}  I_{\Box}^{su}
+ {t u \over s} I_{\Box}^{tu}
\nnl   &
+2 {t^2 + t u + u^2 \over  t u }      I_{\triangle}^s 
+2 {s^2 + s u + u^2 \over  s u }   I_{\triangle}^t 
+2 {t^2 + t s + s^2 \over  t s } I_{\triangle}^u 
+ {\cal R}
\bigg \} 
. \end{align} 

\begin{align}
\label{eq:SUV_0hhphiphi}
\cM[1_h^- 2_h^-3_\phi 4_\phi ]   = &
{ \langle 1 2 \rangle^4 \over \mpl^4 } \bigg \{ 
{ m_\Psi^4 \big[ 8 m_\Psi^2 t u - s (t-u)^2  \big ] \over s^4}  I_{\triangle}^s 
- { 2   m_\Psi^4 (t^2 - 4 t u + u^2 )  \over s^4}  I_{\circ}^s   \bigg \} \nnl &
+{ m_\Psi^2 \langle 1 2 \rangle^4 \over \mpl^2 s^2  }   
\bigg \{ 
2 m_\Psi^2 \big [ (y_s + i y_p)^2 s^2 - 8 y_s^2 m_\Psi^2 \big ]  \big [  I_\Box^{st}   +I_\Box^{tu} +I_\Box^{su}\big ]    
\nnl  \hspace{-2.7cm}  & 
-\frac{t u \left(
32 m_\Psi^2 s  y_s^2
-  (t^2 + u^2)  (y_s + i y_p) ^2
+ 2 t u \left(y_p^2-2 i y_p y_s+3 y_s^2\right)\right)}{s^2} 
I_\Box^{tu}   
\nnl   & 
- {2 t [(y_s + i y_p)^2 s^2 +  8 y_s^2 \big  [  s (t - 2 m_\Psi^2) +  t^2   \big ] \over s^2}      I_\triangle^t
+ 8 y_s^2  {t - u \over s}    I_\circ^t
\nnl   & 
- {2 u [(y_s + i y_p)^2 s^2 +  8 y_s^2 \big  [  s (u - 2 m_\Psi^2) +  u^2   \big ] \over s^2}      I_\triangle^u
- 8 y_s^2  {t - u \over s}    I_\circ^u \nnl &
- \bigg [ c_\Psi (s - 4m_\Psi^2)  + i  \tilde c_\Psi  s \bigg ] I_\triangle^s \bigg \}
+ c_1 I_1 + \mathcal{R} 
\end{align}  

\begin{align}
   \cM[1_\phi 2_\phi 3_\phi 4_\phi] = &
2  \big [   (y_s^2+y_p^2)^2  s t - 32  y_s^4  m_\Psi^4  \big ]  I_\Box^{st} 
+ 2  \big [   (y_s^2+y_p^2)^2  s u - 32  y_s^4  m_\Psi^4  \big ]  I_\Box^{su} 
\nnl  & 
+ 2  \big [   (y_s^2+y_p^2)^2  t u - 32  y_s^4  m_\Psi^4  \big ]  I_\Box^{tu}  +  \bigg \{ 8 \bigg [ 8 c_\Psi m_\Psi^2 y_s^2 + c_\Psi s (y_p^2 - y_s^2)   \nnl &
- 
8 m_\Psi^2 y_s^2 (y_p^2 + y_s^2) - 
s (y_p^4 + 2 \tilde c_\Psi y_p y_s + 2 y_p^2 y_s^2 + y_s^4) \bigg ]  I_\triangle^s 
\nnl &  
+ { c_\Psi^2 (-8 m_\Psi^2 + 2 s) + 
16 c_\Psi m_\Psi^2 (y_p^2 + y_s^2) + 
2 (s \tilde c_\Psi^2 - 4 m_\Psi^2 (y_p^2 + y_s^2)^2) \over m_\Psi^2 }
I_\circ^s \nnl &
+ (s\leftrightarrow t) + (s\leftrightarrow u)
\bigg \} + c_1 I_1 + \mathcal{R}   
. \end{align}  
{\bf  Spin-1}.
\begin{align}
\mathcal{M}[1_h^- 2_h^- 3_h^+ 4_\phi] =& 
{\langle 12 \rangle^4  [3| p_1  p_2  |3]^2  \over  \mpl^3  s t u }  
{  m_V   \over s^2}  \bigg \{   \big [ 6 m_V^4  y_d +  s  ( s - 4 m_V^2) ( y_d + y_a )  \big ] \bigg[ {s t  + 2m_V^2 u \over u}  I_{\Box}^{st} \nnl &
+ {s u  + 2m_V^2 t \over t} I_{\Box}^{su}
+2 \bigg (  {s^2 \over t u  }  - 1 \bigg )   I_{\triangle}^{s} \bigg] 
+ \bigg [ 12 m_V^6 y_d
- { m_V^4 \over s}  \big [ 8s^2 ( y_d + y_a ) - 54 y_d t u  \big ] \nnl &
+ {2 m_V^2 \over s^2} \big [ \big( s^4 + 8 s^3 t \big)( y_d + y_a )+  (26  y_d + 8 y_a ) s^2 t^2 +  36 y_d s t^3 + 18  y_d t^4 
\nnl &  
+ { t  u \over s^3} \big [ \big( s^4 + 4 s^3 t \big) ( y_d + y_a ) + (10 y_d + 4 y_a) s^2 t^2 + 12 y_d s t^3 + 6 y_d t^4 \big ] 
\bigg ]  I_{\Box}^{tu } \nnl &
+{ y_m  m_V^2  \big [ 3 m_V^2 - 2 s \big ]  } \bigg[
{ 2m_V^2 u  + s t \over u}  I_{\Box}^{st}
+ {2m_V^2 t + s u \over t} I_{\Box}^{su}
+2 \bigg (  {s^2 \over t u  }  - 1 \bigg )   I_{\triangle}^{s}\bigg]\nnl &
\hspace{-2 cm}
+y_m\bigg [ 6 m_V^6 
- m_V^4 \left(\frac{27 t^2}{s} + 4 s + 27 t\right ) 
-\frac{2 m_V^2 t u \left(4 s^2+9 s t+9 t^2\right)}{s^2} 
-\frac{t^2 u^2 \left(2 s^2+3 s t+3 t^2\right)}{s^3} \bigg ]  I_{\Box}^{tu }\nnl &
+
\bigg\{\bigg [ \frac{ 12 y_d m_V^4 \left(s^4+3 s t^3+3 t^4\right)}{s t^2 u}
\nnl   &   
-\frac{8 m_V^2 \left((8 y_d + 2 y_a) s^2 t^3+ \big(2 s^3 t^2-s^5\big) (y_d + y_a)+12 y_d s t^4+ 6 y_dt^5\right)}{s^2 t u}
\nnl   &   \hspace{-2.6 cm}
+\frac{2 \left((14 y_d + 8 y_a) s^3 t^3+(22 y_d + 4 y_a) s^2 t^4+ \big(5 s^4 t^2+s^5 t+s^6 \big) (y_d + y_a)+ 18 y_d s t^5+ 6 y_d t^6\right)}{s^3 u}
\bigg ]  I_{\triangle}^{t}
\nnl  &    
+ y_m \bigg [ 
\frac{6 m_V^4 \left(s^4+3 s t^3+3 t^4\right)}{s t^2 u}-\frac{m_V^2 \left(8 s^3 t^2+32 s^2 t^3-4 s^5+48 s t^4+24 t^5\right)}{s^2 t u}\nnl &
+\frac{2 t^2 u \left(2 s^2+3 s t+3 t^2\right)}{s^3}
\bigg ]  I_{\triangle}^{t} \nnl &
+\bigg [ \frac{2 y_d m_V^2 u \left(5 s^2-10 s t+18 t^2\right)}{s t^2}-\nnl &
-\frac{u \left((-6 y_d -8 y_a) s^3+2 (8 y_d + 6 y_a) s^2 t+ 12 y_d s t^2+24 y_d t^3\right)}{2 s^2 t}
\bigg ]  I_{\circ}^{t} \nnl &
+ y_m \bigg [ 
m_V^2 \left(-\frac{5 s^2}{t^2}+\frac{5 s}{t}-\frac{18 t}{s}-8\right)+\frac{3}{2} \left(\frac{4 t^3}{s^2}-\frac{s^2}{t}+\frac{6 t^2}{s}+s+4 t\right)
\bigg ]  I_{\circ}^{t} \nnl  &
 + (t\leftrightarrow u) \bigg \}
+ c_1 I_1 + \mathcal{R}\bigg \} 
. \end{align}
The next amplitude is related to the one with scalar matter running inside the loop as
\begin{align}
    \mathcal{M}[1_h^- 2_h^- 3_h^- 4_\phi]=3(2y_d+y_m)\mathcal{M}^{S=0}[1_h^- 2_h^- 3_h^- 4_\phi].
\end{align}

\begin{align} 
\mathcal{M}^{(2)}[1_h^- 2_h^- 3_\phi 4_\phi] = &
{ m_V^4 \langle 12 \rangle^4  \over 2 \mpl^2 M_*^2  s^2} \bigg \{ 
4  \big [ 
3 m_V^4 (2 y_d +y_m)^2 
- m_V^2 s ( 2 y_d+y_m +2  i y_a  )^2
 \nnl &
+ s^2 (y_d + i y_a)^2  \big ]   
\big[  I_\Box^{st}  +  I_\Box^{su} \big ]  
\nnl & 
+ \bigg [ 12 m_V^2 \left((2 y_d+ y_m)^2 + 4 y_a^2  \right)
-s \left(
8 y_d^2 + 8 y_a^2 +3 y_m^2 
+16 y_d y_m +16 i y_a y_m  \right)  \nnl &
- \frac{4}{m_V^2} c_V  \big [  s^2 - 4 m_V^2 s + 6 m_V^4   \big ]   
-4 i    \tilde c_V    \frac{s}{m_V^2} (s - 4m_V^2)   
+ 2 \hat c_V  {3 m_V^2 (s - 2m_V^2) \over m_V^2} \bigg ]  I_\triangle^s  \nnl &
+ \bigg [  
6 m_V^4 (2 y_d+y_m)^2 + \frac{m_V^2 \left(2 t^2 (2 y_a-i (2 y_d+y_m))^2\right)}{s}
\nnl   &   
+\frac{ m_V^2\left(8 t u \left(2 y_a^2-2 i y_a (2 y_d+y_m)+(2 y_d+y_m)^2\right)+2 u^2 (2 y_a-i (2 y_d+y_m))^2\right)}{s}
\nnl    &   
+ \frac{t^2 u^2 \left(-4 y_a^2+8 i y_a (y_d-y_m)+16 y_d^2+4 y_d y_m+y_m^2\right)}{s^2}\nnl &
+\frac{- t u ( t^2+u^2)  \left(4 y_a^2-8 i y_a y_d+4 i y_a y_m-4 y_d^2+4 y_d y_m+y_m^2\right)}{s^2}
\nnl   \hspace{-2.7cm} &    
-  \frac{2 (t^4 + u^4) (y_a-i y_d)^2 }{s^2}  
\bigg ] I_\Box^{tu}  
+ \bigg [ -\frac{6 t \left(2 m_V^2-t\right) (2 y_d+y_m)^2}{s}\nnl &
+\frac{6 t^3 (2 y_d+y_m)^2}{s^2}-2 t (2 y_a-i (2 y_d+y_m))^2  \bigg ] I_\triangle^t 
\nnl  \hspace{-2.7cm}&  
-\frac{3 (t - u) (2 y_d+y_m)^2}{s}  I_\circ^t
+ (t \leftrightarrow u)\bigg\}
. \end{align}    

\begin{align*}
\cM[1_\phi 2_\phi 3_\phi 4_\phi] &= 
\frac{1}{16 m_V^2 M_*^2} \bigg\{
 16 I_{\Box}^{st} m_V^4 \bigg[ 
-16 s t y_a^4 + \frac{2 s^2 t^2 (y_a^2 + y_d^2)^2}{m_V^4} 
- 16 s t y_a^2 y_d (2 y_d + y_m) \\ &
+ 6 m_V^4 (2 y_d + y_m)^4 - \frac{2 s t (s + t) (y_a^2 + y_d^2)(4 y_a^2 + 4 y_d^2 - y_m^2)}{m_V^2} \\&
+ \frac{1}{2}(s + t)^2 (-4 y_a^2 - 4 y_d^2 + y_m^2)^2 
- s t (2 y_d + y_m)^2 (4 y_d^2 + y_m^2) 
\bigg] 
+ (t \leftrightarrow u) 
\\ & 
+ 16 I_{\Box}^{tu} m_V^4 \bigg[ 
-16 t u y_a^4 + \frac{2 t^2 u^2 (y_a^2 + y_d^2)^2}{m_V^4} 
- 16 t u y_a^2 y_d (2 y_d + y_m) + 6 m_V^4 (2 y_d + y_m)^4 
\\ & 
- \frac{2 t u (t + u)(y_a^2 + y_d^2)(4 y_a^2 + 4 y_d^2 - y_m^2)}{m_V^2} 
+ \frac{1}{2}(t + u)^2 (-4 y_a^2 - 4 y_d^2 + y_m^2)^2 
\\ & 
- t u (2 y_d + y_m)^2 (4 y_d^2 + y_m^2) 
\bigg] 
\\ & 
+ \bigg\{16 I_{\triangle}^{s} m_V^4 \bigg[ 
-64 (t + u) y_a^4 + \frac{8 (t^2 + u^2) y_a^4}{m_V^2} 
- \frac{4 (t^3 + u^3) y_a^4}{m_V^4} 
\\ & 
- 64 \tilde{c}_V (t + u) y_a y_d 
- \frac{16 \tilde{c}_V (t^2 + u^2) y_a y_d}{m_V^2} 
- \frac{32 \tilde{c}_V t u y_a y_d}{m_V^2} 
\\ & 
+ 192 m_V^2 y_a^2 y_d^2 
- 96 (t + u) y_a^2 y_d^2 
+ \frac{16 (t^2 + u^2) y_a^2 y_d^2}{m_V^2} 
- \frac{8 (t^3 + u^3) y_a^2 y_d^2}{m_V^4} 
\\ & 
+ 192 m_V^2 y_d^4 
- 32 (t + u) y_d^4 
+ \frac{8 (t^2 + u^2) y_d^4}{m_V^2} 
- \frac{4 (t^3 + u^3) y_d^4}{m_V^4} 
\\ & 
+ \frac{8 c_V (t + u)^2 (y_a^2 - y_d^2)}{m_V^2} 
- 32 \tilde{c}_V (t + u) y_a y_m 
+ 192 m_V^2 y_a^2 y_d y_m 
- 32 (t + u) y_a^2 y_d y_m 
\\ & 
+ 384 m_V^2 y_d^3 y_m 
- 32 (t + u) y_d^3 y_m 
- \frac{\hat{c}_V (t + u)^2 y_m^2}{m_V^2} 
+ 48 m_V^2 y_a^2 y_m^2 
\\ & 
+ 8 (t + u) y_a^2 y_m^2 
- \frac{4 (t^2 + u^2) y_a^2 y_m^2}{m_V^2} 
- \frac{8 t u y_a^2 y_m^2}{m_V^2} 
+ 288 m_V^2 y_d^2 y_m^2 
\\ & 
- 12 (t + u) y_d^2 y_m^2 
- \frac{4 (t^2 + u^2) y_d^2 y_m^2}{m_V^2} 
- \frac{8 t u y_d^2 y_m^2}{m_V^2} 
+ 96 m_V^2 y_d y_m^3 
\\ & 
- 4 (t + u) y_d y_m^3 
+ 12 m_V^2 y_m^4 
- (t + u) y_m^4 
+ \frac{(t^2 + u^2) y_m^4}{2 m_V^2} 
+ \frac{t u y_m^4}{m_V^2} 
\\ & 
- 24 c_V m_V^2 (2 y_d + y_m)^2 
- 12 \hat{c}_V m_V^2 (2 y_d + y_m)^2 
\\ & 
+ 8 c_V (t + u) (4 y_a^2 - (2 y_d + y_m)^2) 
+ 2 \hat{c}_V (t + u) (8 y_a^2 - (2 y_d + y_m)^2) 
\bigg] 
\\ & 
+ I_{\circ}^{s} \bigg[ 
128 \tilde{c}_V^2 m_V^2 (t + u) 
+ 32 \tilde{c}_V^2 (t^2 + u^2) 
+ 64 \tilde{c}_V^2 t u 
+ 32 c_V^2 (6 m_V^4 + 4 m_V^2 (t + u) + (t + u)^2) 
\\ & 
+ 4 \hat{c}_V^2 (12 m_V^4 + 4 m_V^2 (t + u) + (t + u)^2) 
+ 768 m_V^4 y_a^4 
- 128 m_V^2 (t + u) y_a^4 
+ 32 (t^2 + u^2) y_a^4 
\\ & 
+ 64 t u y_a^4 
+ 1536 m_V^4 y_a^2 y_d^2 
- 256 m_V^2 (t + u) y_a^2 y_d^2 
+ 64 (t^2 + u^2) y_a^2 y_d^2 
\\ & 
+ 128 t u y_a^2 y_d^2 
+ 768 m_V^4 y_d^4 
- 128 m_V^2 (t + u) y_d^4 
+ 32 (t^2 + u^2) y_d^4 
+ 64 t u y_d^4 
\\ & 
- 256 \tilde{c}_V m_V^2 (t + u) y_a y_m 
+ 1536 m_V^4 y_a^2 y_d y_m 
+ 1536 m_V^4 y_d^3 y_m 
\\ & 
+ 384 m_V^4 y_a^2 y_m^2 
+ 96 m_V^2 (t + u) y_a^2 y_m^2 
+ 1152 m_V^4 y_d^2 y_m^2 
+ 96 m_V^2 (t + u) y_d^2 y_m^2 
\\ & 
+ 384 m_V^4 y_d y_m^3 
+ 32 m_V^2 (t + u) y_d y_m^3 
+ 48 m_V^4 y_m^4 
+ t^2 y_m^4 + 2 t u y_m^4 + u^2 y_m^4 
\\ & 
+ 16 c_V m_V^2 \bigg( 
6 \hat{c}_V (2 m_V^2 + t + u) 
- (t + u) (8 y_a^2 + 8 y_d^2 + 16 y_d y_m + 3 y_m^2) 
\\ & 
- 12 m_V^2 (4 y_a^2 + (2 y_d + y_m)^2) 
\bigg) 
- 4 \hat{c}_V \bigg( 
(t + u)^2 y_m^2 
+ 2 m_V^2 (t + u) y_m (8 y_d + y_m) 
\\ & 
+ 24 m_V^4 (4 y_a^2 + (2 y_d + y_m)^2) 
\bigg) 
\bigg] \bigg\} + (t \leftrightarrow u)
\bigg\}  + c_1 I_1 + \mathcal{R}.
\end{align*}

\newpage
\bibliographystyle{JHEP}
\bibliography{gaussbonnet}

\end{document}